\documentclass[%
 reprint,
 superscriptaddress,
 amsmath,amssymb,
 aps,
 prc,
]{revtex4-2}

\usepackage{graphicx}
\usepackage{dcolumn}
\usepackage{bm}
\usepackage[colorlinks=true, pdfstartview=FitV, linkcolor=red, citecolor=blue, urlcolor=blue]{hyperref}

\usepackage{xcolor}

\begin{document}


\title{$\Lambda$ polarization from vortex ring as medium response for jet thermalization}

\author{Vítor Hugo Ribeiro}%
\email{vribeiro@ifi.unicamp.br}
\affiliation{Instituto de Física Gleb Wataghin, Universidade Estadual de Campinas, Campinas, Brasil}%

\author{David Dobrigkeit Chinellato}%
\email{daviddc@unicamp.br}
\affiliation{Instituto de Física Gleb Wataghin, Universidade Estadual de Campinas, Campinas, Brasil}%

\author{Michael Annan Lisa}
\email{lisa@physics.osu.edu}
\affiliation{The Ohio State University, Columbus, Ohio, USA}

\author{Willian Matioli Serenone}
\email{wmatioli@if.usp.br}
\affiliation{Instituto de Física da USP, Universidade de São Paulo, São Paulo, Brasil}
\affiliation{Department of Physics, University of Illinois at Urbana-Champaign, Urbana, IL 61801, USA}%

\author{Chun Shen}
\email{chunshen@wayne.edu}
\affiliation{Department of Physics and Astronomy, Wayne State University, Detroit, MI 48201, USA}%
\affiliation{RIKEN BNL Research Center, Brookhaven National Laboratory Upton, NY 11973, USA}

\author{Jun Takahashi}%
\email{jun@ifi.unicamp.br}
\affiliation{Instituto de Física Gleb Wataghin, Universidade Estadual de Campinas, Campinas, Brasil}%

\author{Giorgio Torrieri}%
\email{torrieri@unicamp.br}
\affiliation{Instituto de Física Gleb Wataghin, Universidade Estadual de Campinas, Campinas, Brasil}%

\date{\today}
             
\begin{abstract}
We performed a systematic study on the formation of vorticity rings as the process for jet thermalization in the medium created in high-energy nuclear collisions. In this work, we expanded our previous analysis to a more realistic framework by considering non-central events and fluctuations in the initial condition. We simulate the formation and evolution of the flow vortex structure in a relativistic viscous hydrodynamic model and study the sensitivity of the proposed ``ring observable'' ($\mathcal{R}^{t}_{\Lambda}$) that can be measured experimentally through the polarization of $\Lambda$ hyperons. We show that this observable is robust with respect to fluctuating initial conditions to capture the jet-induced vortex flow signal and further study its dependence on different model parameters, such as the jet's velocity, position, the fluid's shear viscosity, and the collision centrality. The proposed observable is associated with the formation of vorticity in a quark-gluon plasma, showing that the measurement of particle polarization can be a powerful tool to probe different properties of jet-medium interactions and to understand better the polarization induced by the transverse and longitudinal expansions of the medium.
\end{abstract}

\maketitle


\section{\label{sec:Introduction} Introduction}

In a previous work \cite{Serenone2021}, we showed that the hydrodynamic nature \cite{Sch_fer_2009, GALE2013, Derradi2016, Shuryak2017, Werner_2012, shen2020recent, Nunes_da_Silva_2023} of the strongly interacting matter created in a heavy-ion collision presents a unique opportunity to study one of the most intriguing topics in the field, which is the jet quenching phenomenon \cite{GYULASSY1990, Blaizot_2015, Wang1992, Wiedemann_2010, Cao_2021, Aad2010, Chatrchyan2011}. The main goal of the experiments that measure high-energy heavy-ion collisions, such as those performed at LHC and RHIC, is to probe the QCD phase diagram and detect novel properties of the nuclear matter in extreme temperature conditions. Based on this search, we developed a model that utilizes the fluid description of the Quark-Gluon Plasma (QGP) and the coupling between vorticity and polarization \cite{vort_pol_lisa, STAR}  to investigate the thermalization of a quenched jet within the hydrodynamic medium.

In this model, we assumed jet thermalization in the medium. We demonstrated that the energy and momentum deposited from the jet would generate a typical hydrodynamic structure: the vorticity ring. We also showed that such a structure would survive the explosive hydrodynamic evolution and that the effects of the ring will ultimately lead to a polarization pattern of the $\Lambda$ hyperons emitted by the system. To measure this polarization pattern, we proposed an observable called the ``ring observable'',  $\mathcal{R}^{t}_{\Lambda}$~\cite{Lisa2021}. In the present work, we expand on the previous analysis by considering non-central events and fluctuating initial conditions. We systematically study the thermalization of the energy lost by the jet through the vorticity ring phenomenon, evaluating its dependence on different model parameters related to hydrodynamic evolution. We also provide a general picture of how it can be experimentally observed in the polarization of $\Lambda$ hyperons. As a brief extension of this work, we use $\mathcal{R}^{t}_{\Lambda}$ to evaluate the polarization induced by the expansion of the hydrodynamic system and show that the ring observable, in addition to detecting the effects generated by jet thermalization, can also be used to decouple the contributions of polarization due to transversal and longitudinal dynamics.

\section{\label{sec:Methodology} Methodology}

\subsection{Simulation Workflow}

The current study is performed with a framework upgraded from our previous analysis \cite{Serenone2021}. In particular, we have improved the method of inserting the jet's energy-momentum current into the medium. We can now control the position and time of energy-momentum deposition and the direction of the momentum of the thermalized jet. 
Also, to test the robustness of the proposed observable, we have utilized more realistic fluctuating initial conditions in our simulations.

\begin{figure}[]
    \includegraphics[scale=0.16]{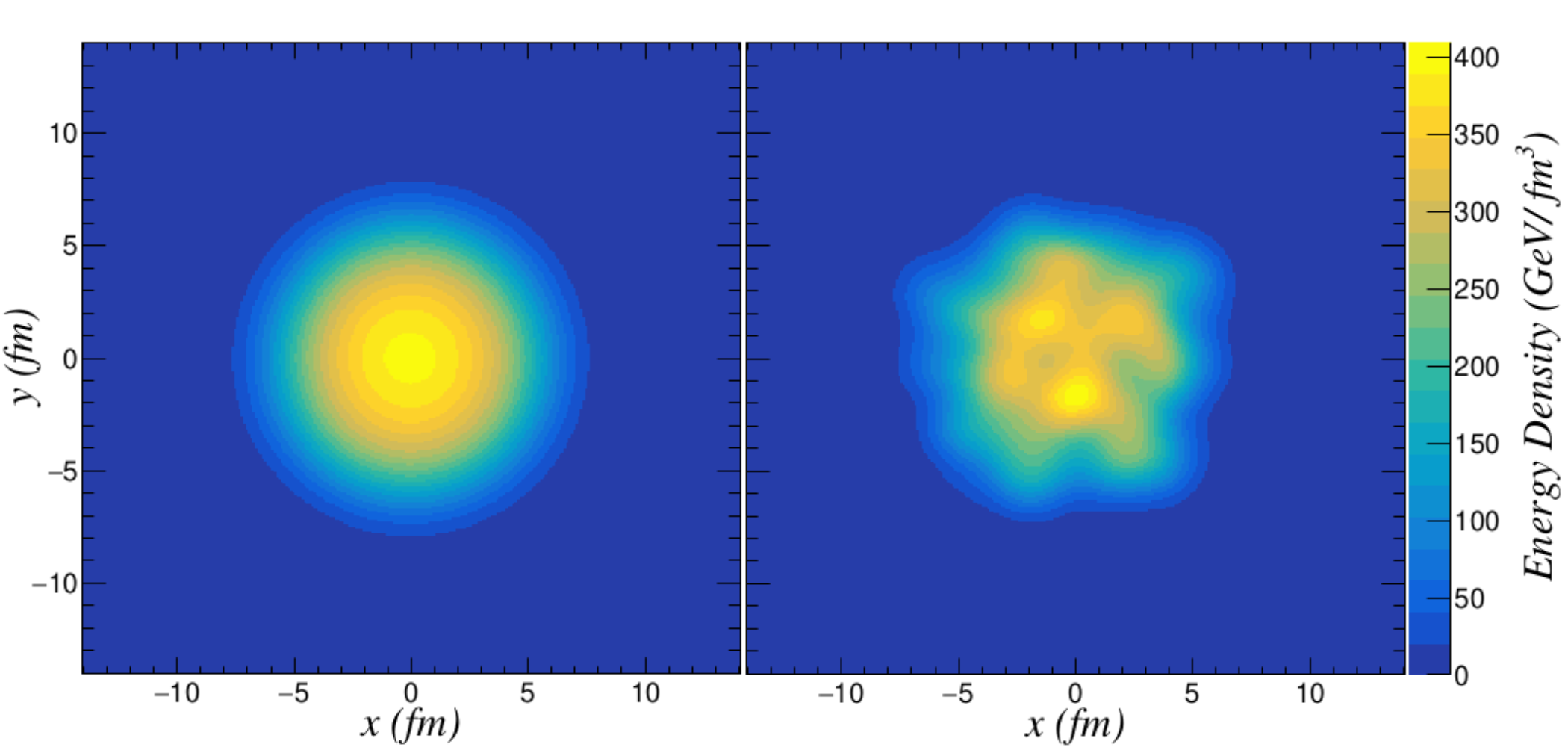}
    \caption{\label{fig:ICs} Mid-rapidity energy density distribution (at $\eta_s = 0$) generated by the 3D $\text{T}_{\text{R}}\text{ENTo}$ model \cite{trento} for the initial condition (IC) profiles of Pb+Pb collisions at $2.76$ TeV with 0-5\% centrality class.
    The left panel represents an IC created from an average of 10,000 fluctuating initial conditions, while the right panel shows a typical fluctuating initial condition.}
\end{figure}

To generate the initial conditions (IC) for our simulations, we utilized the $\text{T}_{\text{R}}\text{ENTo}$ 3D \cite{trento} model. It provides initial-state energy density profiles event-by-event for Pb-Pb collisions at $\sqrt{s_{NN}}= 2.76$ TeV. We used the same parameters as in our previous work \cite{trento, bernhard2018} to generate the initial conditions for the system. The parameters common to both 2D and 3D $\text{T}_{\text{R}}\text{ENTo}$ were obtained from Ref.~\cite{trento}, while those exclusive to 3D $\text{T}_{\text{R}}\text{ENTo}$ were taken from Ref.~\cite{bernhard2018}. Table \ref{tab:trento-parameters} explicitly shows the main parameters used to generate the ICs. We do not consider sub-nucleonic structures in this work.
The energy distributions of each IC were also normalized by matching the full simulation results with the mid-rapidity charged particle multiplicity measured from central (0-5\%) events by the ALICE experiment \cite{alice_data}.

\begin{table}[h]
\caption{\label{tab:table1} \label{tab:trento-parameters} List of parameters used in $\text{T}_{\text{R}}\text{ENTo}$ 3D to generate the initial conditions.}
\begin{ruledtabular}
\begin{tabular}{ccc}
 Parameter & Description & Value  \\
\colrule
$p$ & Reduced Thickness    & $0.007$  \\
$w$ &  Gaussian nucleon width   & $0.956$ fm \\
 & Skewness type & Relative Skewness            \\
$\mu_0$ & Rapidity shift mean coeff. & $0.0$ \\
$\sigma_0$ & Rapidity width std. coeff. & $2.9$ \\
$\gamma_0$ & Rapidity skewness coeff. & $7.3$ \\
$J$ & Pseudorapidity Jacobian param. & $0.75$\\
$d_{min}$ & Nucleon minimum distance & $1.27$ fm \\
\end{tabular}
\end{ruledtabular}
\end{table}

For each centrality bin, we sample multiple impact parameters $b$ to generate initial conditions as listed in Table \ref{tab:centrality}. The intervals presented were defined using a minimum bias analysis with one million initial conditions, which were ordered according to the total initial energy of each IC provided by $\text{T}_{\text{R}}\text{ENTo}$ 3D.
A comparison between the initial condition produced with the method used in the previous work \cite{Serenone2021} and the current one is shown in Fig.~\ref{fig:ICs}.
As can be seen, the left panel shows the profile of a smooth initial condition obtained from an average of 10,000 fluctuating events by aligning impact parameters along the $+x$-axis. The resulting initial condition captures the average geometry for each centrality bin.
In contrast, the right panel shows one of the initial conditions used in this new event-by-event analysis.

\begin{table}[h]
\caption{\label{tab:table2} \label{tab:centrality} Impact parameter intervals used to define each centrality class.}
\begin{ruledtabular}
\begin{tabular}{cccc}
 & $b$ (fm) & Centrality (\%) &  \\
\colrule
& $0.0 - 3.74$    & 0 - 5    &    \\
& $7.46 - 9.13$   & 20 - 30  &    \\
& $10.55 - 11.79$ & 40 - 50  &    \\
& $12.91 - 13.94$ & 60 - 70  &    \\
\end{tabular}
\end{ruledtabular}
\end{table}

In the next stage, the initial conditions are hydrodynamically evolved using a viscous relativistic theory solved by MUSIC code \cite{music1,music2,music3}. MUSIC performs a (3+1)D evolution of the system, which is required for vorticity calculations. We use a grid spacing equal to $dx = dy = 0.1$ fm in the $x-y$ plane and $d\eta_s = 0.2$ in the space-time rapidity $\eta_s$ direction to ensure numerical accuracy.
In this stage, we applied the lattice-QCD equation of state from the HotQCD Collaboration \cite{hotQCDeos} and started the evolution at a longitudinal proper time $\tau_0 = 0.25$ fm/$c$. We neglect the net baryon density evolution at LHC for the hydrodynamic evolution. We also keep the specific shear viscosity of the medium in a constant value of $\eta/s = 0.08$ for all results, except for those expressed in Section~\ref{sec:Results-Viscosity-Scan}. Bulk viscous effects are not included.
The system then evolves until every cell of the grid reaches a freeze-out temperature of $T = 151$ MeV.

During the hydrodynamic stage, in order to simulate the dynamics of the energy absorbed by the medium from the quenched jet, a hot source term with well-defined energy and momentum is deposited into the system.
Physically, this model can be addressed to a scenario in which we have a $\gamma$-jet event inside the fluid, where only the jet will interact with the QGP and $\gamma$ would provide a trigger direction for the analysis. 
In contrast to our previous work, the quenched jet's energy-momentum currents are deposited directly during the hydrodynamic evolution at a fixed time $\tau_\mathrm{th} = 1.0$ fm/$c$. The parameters that characterize the hot source term inserted into the hydrodynamic evolution are kept the same as in Ref.~\cite{Serenone2021}, with $E_{th} = 59.6$ GeV, $p_{th} = 43.0$ GeV/$c$ and $R_{xy} = 1.0$ fm (radius of the hot source term in the $x-y$ plane), $R_{\eta} = 0.4$ (radius on the longitudinal direction). We will vary some of those parameters to perform a systematic analysis of the dependence of the polarization on them.

After the hydrodynamic phase, fluid cells are converted to particles via the Cooper-Frye formalism \cite{Cooper1974, Cooper1975} through the iSS Sampler code \cite{iSS}. The polarization of the primary emitted $\Lambda$ hyperons is computed on the hydrodynamic hyper-surface \cite{vort_pol_lisa, Becattini_2015}.

\subsection{\label{sec:Methodology-Polarization} Vorticity, Polarization and the Ring Observable}

To compute the polarization, we calculate the spin vector of the $\Lambda$ hyperons when they emit from the particlization hyper-surface at $T = 151$\,MeV. We use the thermal vorticity defined as~\cite{ther-vort1, ther-vort2}, 
\begin{equation}
    \varpi^{\mu \nu} = -\dfrac{1}{2}\left[ \partial^{\mu}(u^{\nu}/T) - \partial^{\nu}(u^{\mu}/T) \right],
\end{equation}
to calculate the $\Lambda$'s polarization from each fluid cell of the freeze-out hypersurface \cite{vort_pol_lisa, Becattini_2015},
\begin{equation}\label{eq:mean_spin}
    S^{\mu}(p) = -\dfrac{1}{8m}\epsilon^{\mu \rho \sigma \tau} p_{\tau}\dfrac{\int d\Sigma_{\lambda}p^{\lambda}n_{F}(1-n_{F})\varpi _{\rho \sigma}}{\int d\Sigma_{\lambda}p^{\lambda}n_{F}}.
\end{equation}
The polarization vector of $\Lambda$ hyperons can be computed as
\begin{equation}
    P^\mu_{\Lambda}(p) = \frac{S^\mu(p)}{\langle S \rangle}.
\end{equation}
with $\langle S \rangle = 1/2$.

\begin{figure*}[t]
\includegraphics[scale = 0.32]{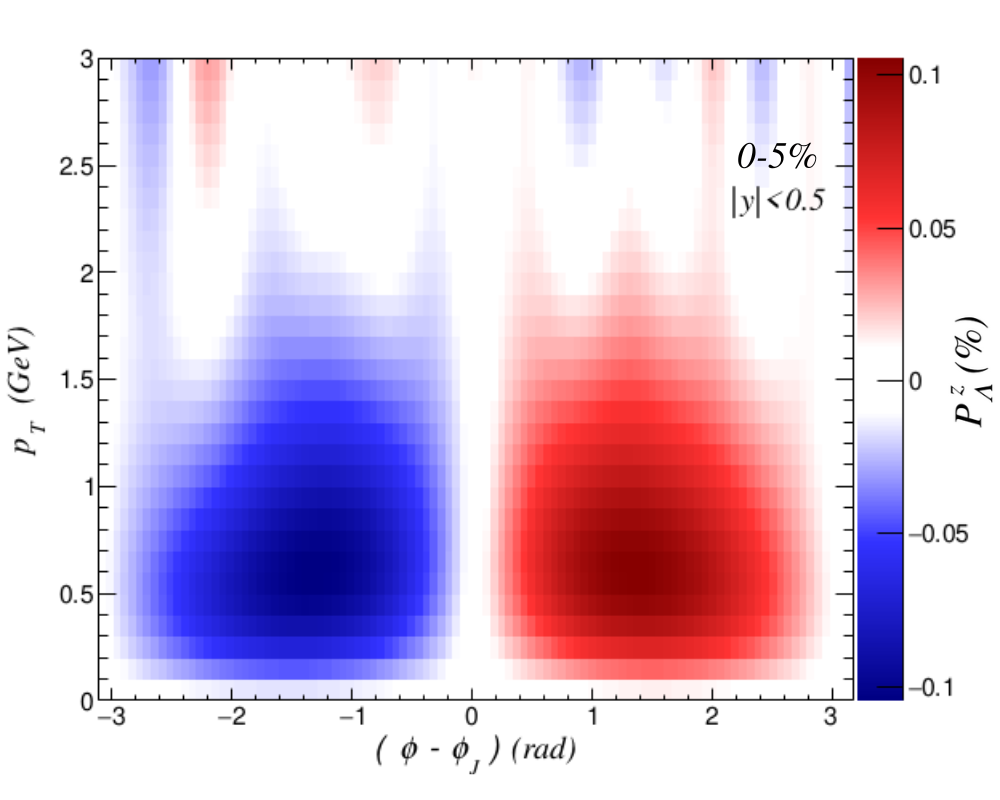}
\quad
\includegraphics[scale = 0.32]{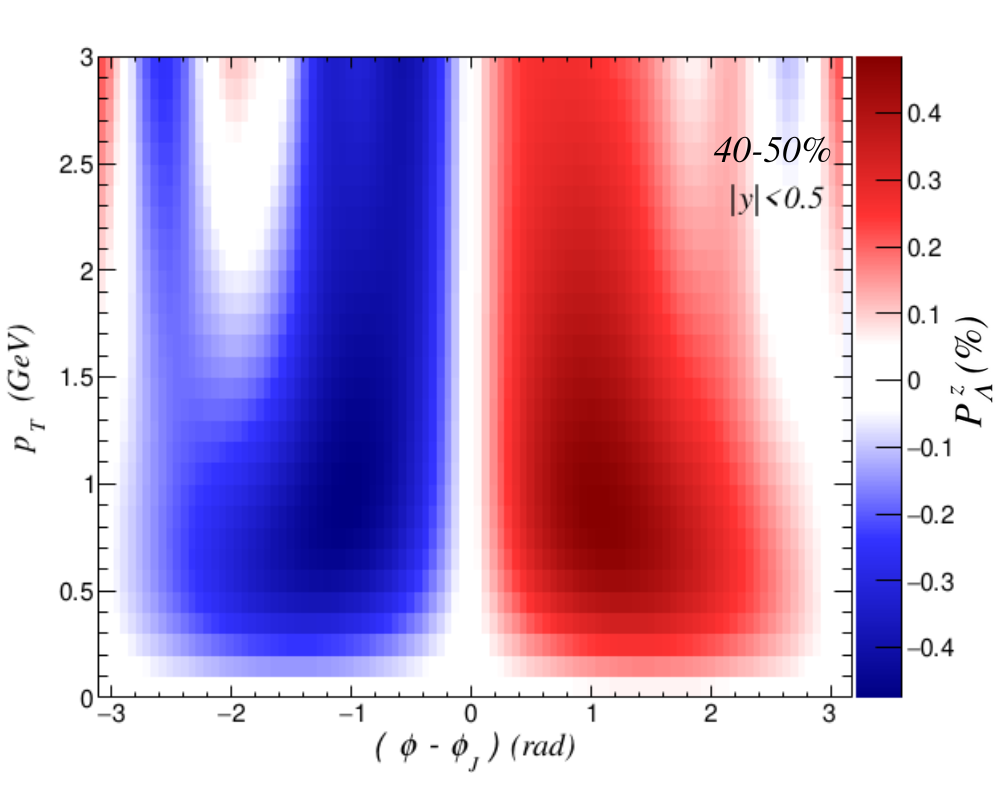}
\caption{\label{fig:sz-color-map} The $\Lambda$'s longitudinal polarization $P^{z}_{\Lambda}$ as a function of transversal momentum $p_{T}$ versus relative azimuthal angle ($\phi - \phi_{J}$) for events of centrality (left) $0-5 \%$ and (right) $40-50\%$. In both centrality bins, we used fluctuating initial conditions, and the jet energy and momentum were deposited at position $x=y=0$, with the direction being randomly selected from a range of $0<\phi_{J}<2\pi$.}
\end{figure*}

To study the $\Lambda$ polarization induced by the vortex ring, we follow the same analysis procedure used in our previous work \cite{Serenone2021}.
Our analysis begins by examining the dependence of the $z$-component of the $\Lambda$ polarization vector in the lab frame on the transverse momentum and the azimuthal angle relative to the direction of the quenched jet.
This result is presented in Fig.~\ref{fig:sz-color-map} for two different centrality classes, $0-5 \%$ (left) and $40-50\%$ (right) in Pb+Pb collisions. For this analysis, the jet energy and momentum were deposited at the center of the fireball $x=y=\eta_s=0$, and the azimuthal direction of propagation was selected from a random uniform distribution using the interval $[0, 2\pi]$ with rapidity $y = 0$.
The color gradient indicates that in both centrality bins, the longitudinal polarization vector $P_{\Lambda}^{z}$ is concentrated in a region with $p_{T}<1.5$ GeV, and its magnitude reaches a maximum around $\phi - \phi_{J} \sim 1$ rad.
Additionally, it is possible to see some small signals in $p_{T} > 2.5$ GeV$/c$, which oscillates with respect to the angle axis.
Those residual signals were attributed to a statistical fluctuation from the polarization induced by the anisotropic expansion of the fluid, which survived the randomization method applied in the direction of the momentum of the thermalized jet. Both the signal and the method will be discussed later in this section. 
The comparison between the polarization observed in central and peripheral events, as shown in Fig.~\ref{fig:sz-color-map}, reveals that the distribution of polarization reaches bigger values in peripheral events. Such an observation will be discussed in more detail in Section \ref{sec:Results-Position-Centrality-Scan}.

\begin{figure*}[t]
\includegraphics[scale = 0.34]{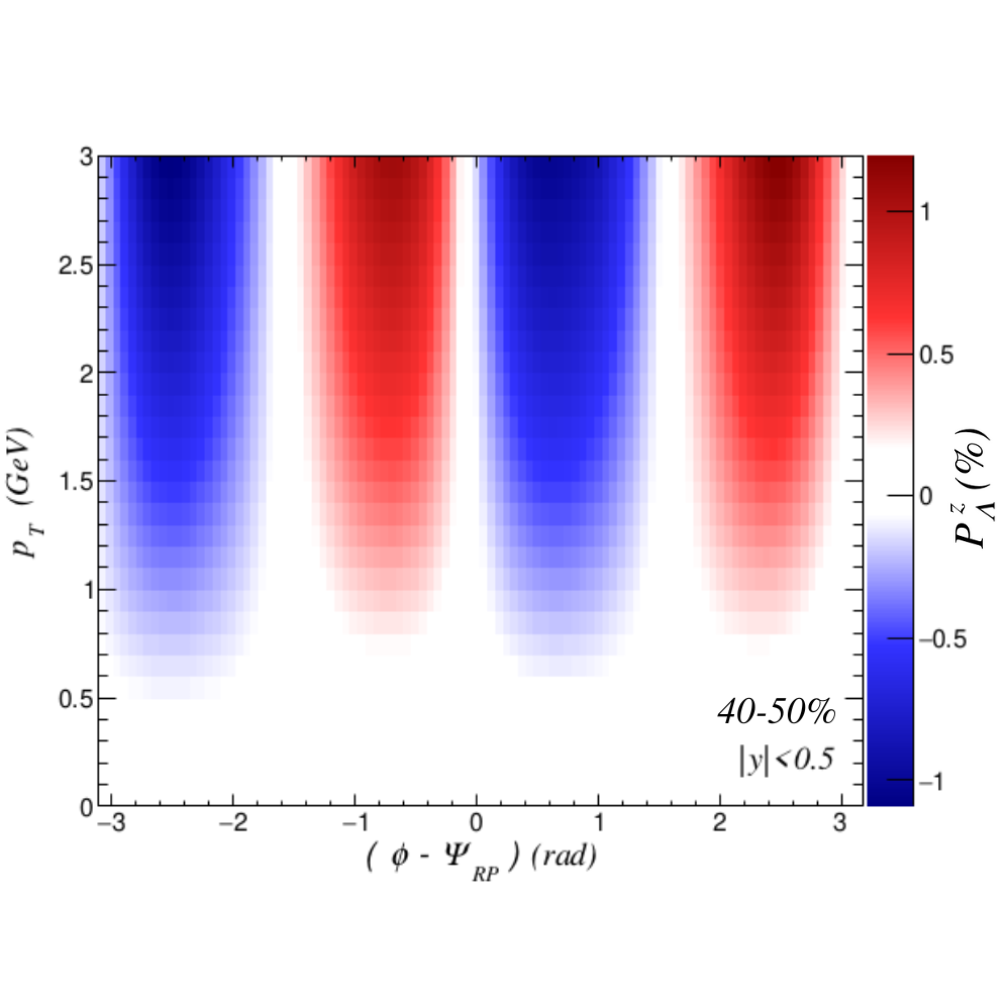}
\quad
\includegraphics[scale = 0.3]{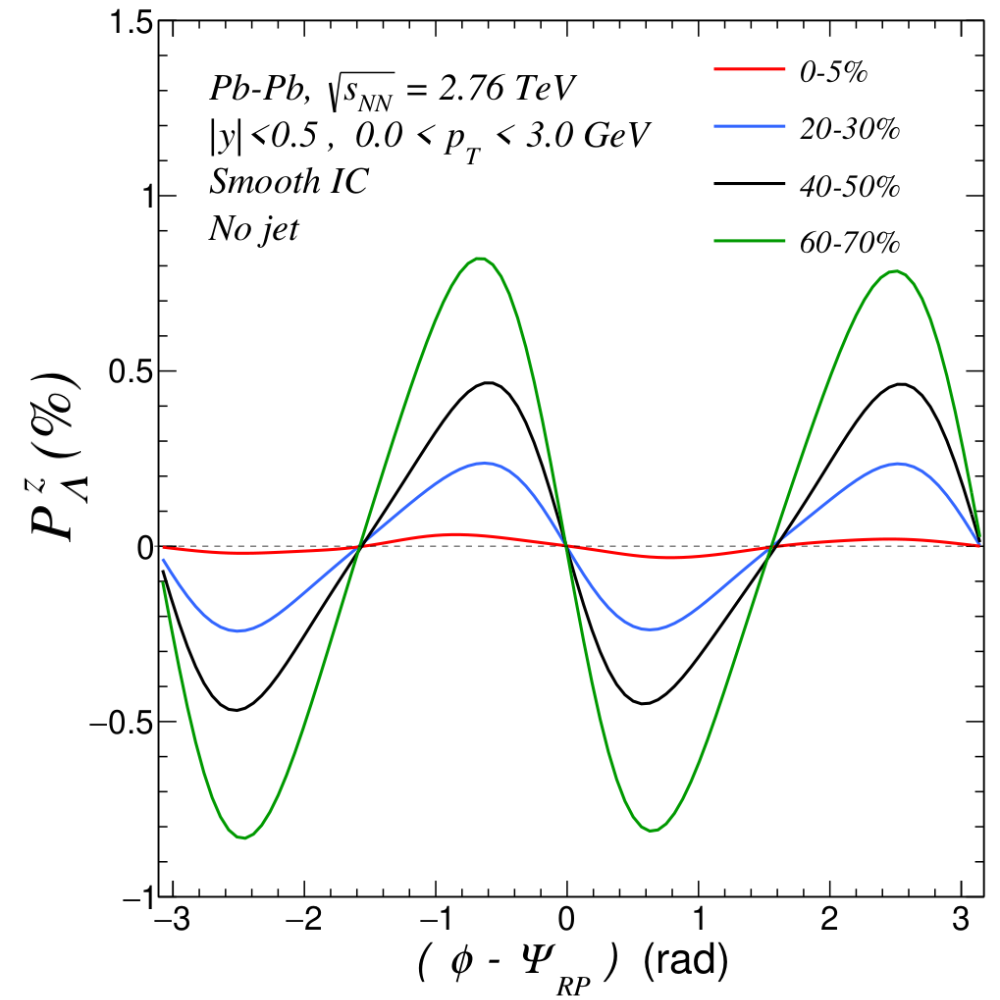}
\caption{\label{fig:sz-no-jet} (Left) The $\Lambda$'s longitudinal polarization $P^{z}_{\Lambda}$ as a function of transversal momentum $p_{T}$ and relative azimuthal angle ($\phi - \phi_{J}$) for events of centrality $40-50\%$. This event-by-event analysis was made considering fluctuating initial conditions, and no thermalized jet was inserted during the hydrodynamics evolution. (Right) Polarization $P_{\Lambda}^{z}$ as a function of the azimuthal angle for different collision centralities obtained from simulations with smooth initial conditions. In both plots, the angle $\Psi_{RP}$ represents the orientation of the reaction plane.}
\end{figure*}

In heavy-ion collisions, local vorticity can originate from different sources. We consider two sources in our work: the gradients created by the energy-momentum currents deposited by the quenched jet and the gradients of initial-state hot spots which drive the collective expansion of the QGP fluid \cite{vort_collect1, vort_collect2, vort_collect3, vort_collect4}. 
To evaluate the magnitude and distribution of the background polarization generated by the medium's collective expansion, we studied $\Lambda$'s polarization without the energy-momentum deposition from jets into the medium in Fig.~\ref{fig:sz-no-jet}.
In this case, the $\Lambda$'s polarization appears to be dominated by the pressure gradient anisotropy and reflects the effects of elliptic flow $v_{2}$.
The left panel of Fig.~\ref{fig:sz-no-jet} shows the lab frame $\Lambda$'s longitudinal polarization ($P_{\Lambda}^{z}$) as a function of the relative azimuthal angle between $\Lambda$'s momentum and the reaction plane ($\phi-\Psi_{RP}$) for 40-50\% Pb+Pb collisions.
The $p_T$-dependence of $P_{\Lambda}^{z}$ is qualitatively different compared to those induced from quenched jets as shown in Fig.~\ref{fig:sz-color-map}. The elliptic flow-induced longitudinal polarization $P_{\Lambda}^{z}$ increases with $p_T$ and has a characteristic quadripolar pattern in the transverse plane. The same behavior was observed for the 0-5\% centrality class.
The right panel of Fig.~\ref{fig:sz-no-jet} shows the dependency of the longitudinal polarization as a function of collision centrality. The magnitude of oscillation increases with collision centrality together with the size of the elliptic flow in the system. Comparing Figs.~\ref{fig:sz-color-map} and \ref{fig:sz-no-jet}, we find that the polarization pattern induced by the quenched jets does not wash out by the elliptic flow. Because the jet's direction is random with respect to the event plane of elliptic flow, the background polarization that appears in Fig.~\ref{fig:sz-no-jet} is averaged out in Fig.~\ref{fig:sz-color-map}.

The vortical polarization pattern induced by the quenched jet can be quantified through the ring observable $\mathcal{R}^{t}_{\Lambda}$ presented in Ref.~\cite{Lisa2021}. 
This proposed experimental observable was developed as a tool to isolate the circular pattern of $\Lambda$ polarization that was induced by the vorticity created in the energy-momentum current deposition by the quenched jets. 
To calculate $\mathcal{R}^{t}_{\Lambda}$, we use the polarization of $\Lambda$'s measured in the lab frame and apply the following formula,
\begin{equation}\label{ring_observable}
    \mathcal{R}^{t}_{\Lambda} \equiv \left \langle \dfrac{\Vec{P}_{\Lambda} \cdot (\hat{t} \times \Vec{p}_{\Lambda})}{|\hat{t} \times \Vec{p}_ {\Lambda}|} \right \rangle_{p_{T}, y},
\end{equation}
where we choose $\hat{t} = \hat{J}$ with $\hat{J}$ corresponding to the propagation direction of the energy-momentum current from the quenched jet. In Eq.~\eqref{ring_observable}, we use $\Lambda$'s multiplicity as the weight to evaluate the average denoted by $\langle \cdots \rangle_{p_{T}, y}$. 
The ring observable are calculated in the kinematic range of $0.5 \: \text{GeV/}c < p_{T} < 1.5 \: \text{GeV/}c$ and $|y|<0.5$. We chose this $p_{T}$ range to focus on the region that appears to be most affected by the thermalization of the jet shown in Fig.~\ref{fig:sz-color-map}. The upper limit was selected to minimize the contributions that could be influenced by the anisotropic expansion, while the lower limit, in addition to accounting for experimental acceptance, also represents the point at which the weighted average will contribute significantly to the final signal, since the multiplicity of $\Lambda$ hyperons in the region of $p_{T}<0.5$ GeV$/c$ is small. For the rapidity interval, it was selected to match the usual acceptance of the experiments.

\section{\label{sec:Results} Results}

\subsection{Influence of initial state fluctuations on the ring observable}

\begin{figure}[h!]
\includegraphics[scale = 0.32]{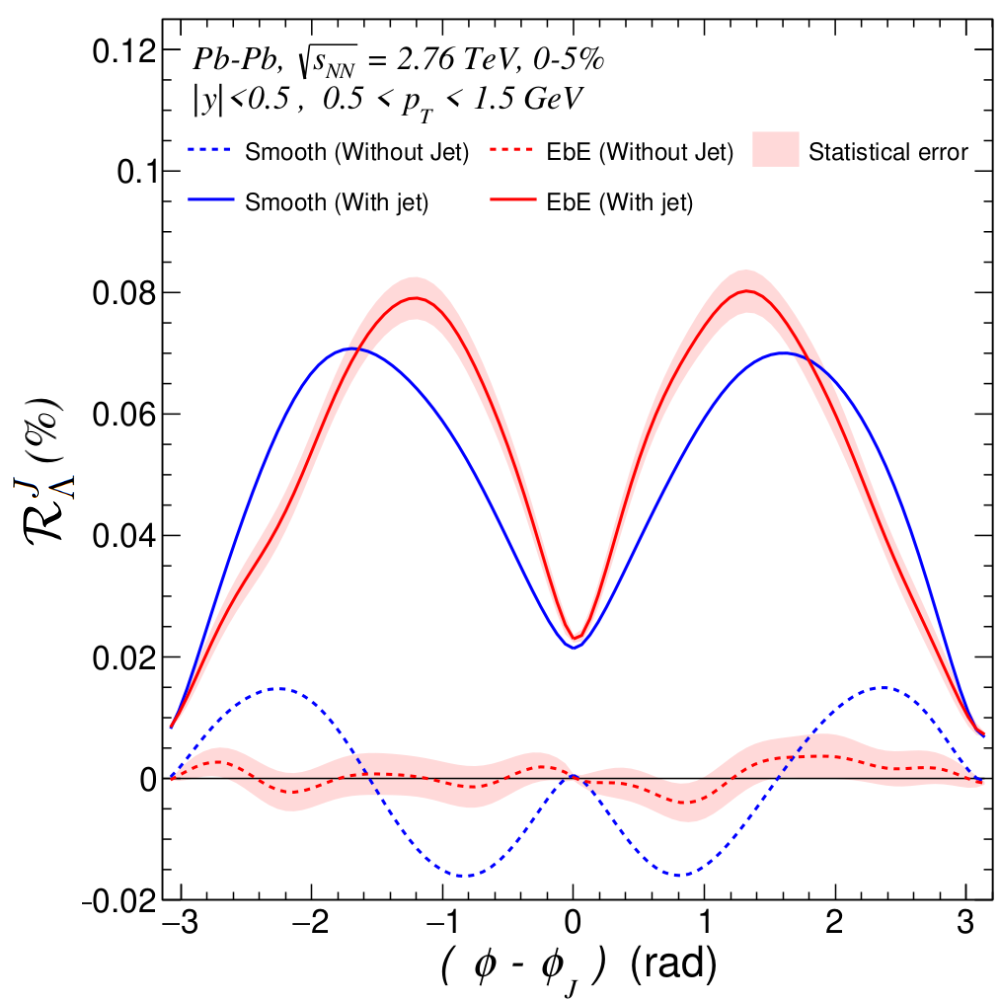}
\caption{\label{fig:smooth-vs-ebe} The ring observable $\mathcal{R}^{J}_{\Lambda}$ as a function of the relative azimuthal angle ($\phi - \phi_{J}$) for event-by-event and smooth initial conditions and with and without the energy-momentum deposition from thermalized jets. Solid curves show the results obtained from simulations in which the energy-momentum currents from quenched jets are deposited. In contrast, the dashed curves correspond to results without jets in the simulations. The red curves correspond to the results of an event-by-event analysis performed with fluctuating initial conditions, while the blue curve corresponds to the results of an analysis done with just one event with a smooth initial condition.}
\end{figure}

In Fig.~\ref{fig:smooth-vs-ebe}, we compare the results obtained from the event-by-event simulations and those from an event-averaged smooth initial condition. On the one hand, the event-by-event results were obtained from 250 fluctuating initial conditions events with impact parameters sampled in the 0-5$\%$ centrality class. In this analysis, the azimuthal direction of the quenched jet $\phi_{J}$ in each event is randomly selected between 0 and $2\pi$. On the other hand, in the smooth event analysis, one smooth initial condition was generated by averaging 10,000 fluctuating ICs.
This event-averaged initial profile has a small but non-zero initial eccentricity, equivalent to the averaged eccentricity in 0-5$\%$ Pb+Pb collisions. In the latter case, the energy-momentum current from the quenched jet was deposited in the direction of the reaction plane, which corresponds to similar conditions as the analysis done in the previous work~\cite{Serenone2021}. In both cases, the energy-momentum current of the quenched jet was deposited at the center of the fireball $x=y=\eta_s=0$. 
The solid lines in Fig.~\ref{fig:smooth-vs-ebe} correspond to the results obtained when the energy-momentum current is deposited, and the dashed lines to the scenario where no source terms from the jet were inserted.
We set $\phi_{J} = \Psi_{RP}$ in the smooth IC case without source terms from the quenched jet. As discussed earlier, we see a small but non-vanishing signal generated by the anisotropic flow.
In the event-by-event scenario without jets, since the jet direction and the reference direction (in the case of no jet) were randomized, we see an almost vanishing signal for $\mathcal{R}_\Lambda^J$, which is expected from our previous discussion on background polarization.
Regarding the signal generated by the quenched jets, we see that the smooth initial condition scenario is qualitatively similar to the more realistic event-by-event simulations. The quantitative difference between both scenarios is attributed to the influence of the transverse expansion. Since the signal obtained with the smooth initial condition was not averaged over different trigger directions for the thermalized jet, the resulting signal (blue solid line) will present a great contribution to the background (blue dashed line).

\subsection{ \label{sec:Results-Viscosity-Scan} Shear viscous effects on the ring observable}

\begin{figure}[]
\includegraphics[scale = 0.32]{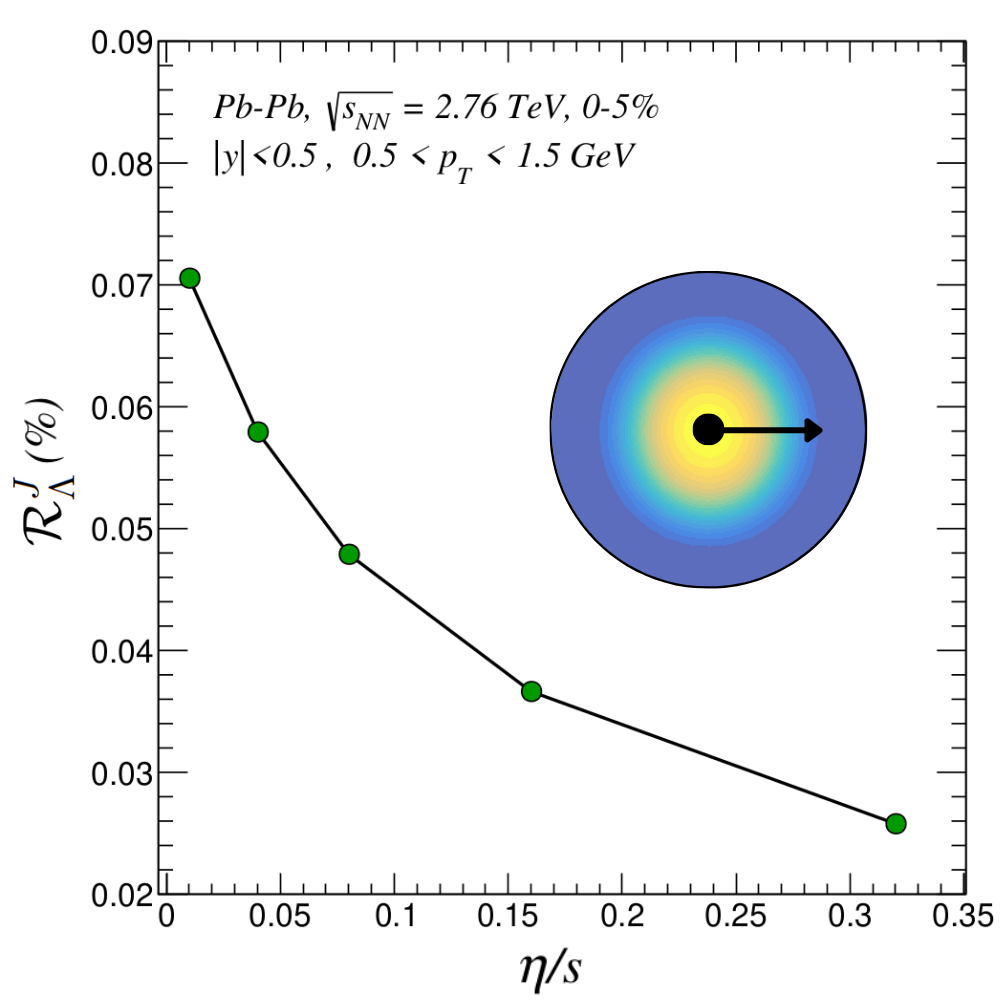}
\caption{\label{fig:visc_scan} Integrated ring observable $\mathcal{R}^{J}_{\Lambda}$ as a function of specific shear viscosity of the medium. The illustration inside the plot represents schematically the scenario studied, in which we applied a smooth initial condition and inserted the energy-momentum current from the quenched jet at the center of the event with momentum pointing in the positive $x$ axis.}
\end{figure}

In this section, we investigate the influence of different values for the specific shear viscosity $\eta/s$ of the medium on the jet-induced ring observable $\mathcal{R}_\Lambda^J$. Based on the comparisons shown in Fig.~\ref{fig:smooth-vs-ebe}, simulations with event-averaged initial conditions can provide a good estimate for the ring observable. Therefore, we systematically scan through different specific shear viscosity values with the event-averaged smooth IC. For these simulations, the energy-momentum current from the quenched jet was deposited to the hydrodynamic medium at $x=y=0$, with the momentum direction pointing to the positive $x$ direction. The integrated ring observable $\mathcal{R}_\Lambda^J$ results are presented in Fig.~\ref{fig:visc_scan}. The magnitude of the $\mathcal{R}_\Lambda^J$ monotonically decreases with increasing the specific shear viscosity of the medium. These results are in qualitative agreement with those in our previous work \cite{Serenone2021}.
Our result indicates that increasing the viscosity of the medium leads to a reduction in the polarization induced by the thermalization of the jet energy. A larger viscosity value prevents the system from developing the velocity gradients required to generate the vorticity ring, leading to a weaker vorticity field and, in turn, a weaker final hadron polarization.

\subsection{The ring observable's sensitivity on jet's velocity}

\begin{figure}[]
\includegraphics[scale = 0.32]{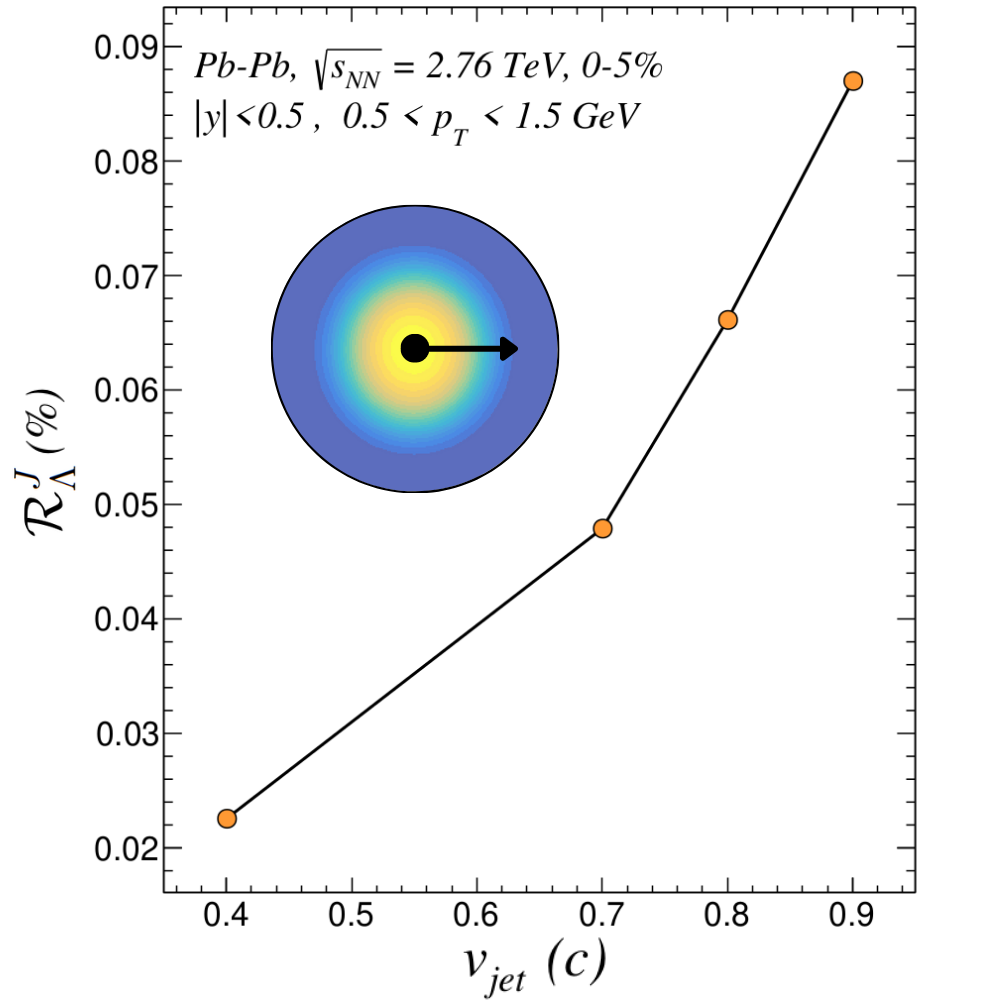}
\caption{\label{fig:momentum_scan} Integrated $\mathcal{R}^{J}_{\Lambda}$ for different jet velocities. The illustration inside the plot represents schematically the scenario studied, in which we applied a smooth initial condition and inserted the jet at the center of the event with momentum pointing in the $+x$ axis.}
\end{figure}

In Fig.~\ref{fig:momentum_scan} we explore the dependence of the ring observable $\mathcal{R}^{J}_{\Lambda}$ on the jet velocity $v_\mathrm{jet} = p_\mathrm{jet}/E_\mathrm{jet} = 0.4c$, $0.7c$, $0.8c$, and $0.9c$ along the $+x$ direction. In this exercise, we fixed the mass of the jet $E_\mathrm{jet} = 59.6$ GeV and calculated its momentum $p_\mathrm{jet}$ according to each velocity value.
To save simulation time, we perform these calculations with an event-averaged IC for 0-5\% Pb+Pb collisions at 2.76 TeV with the jet energy-momentum inserted at position $x=y=0$, and the velocity always pointing in the $+x$ direction. 
Figure~\ref{fig:momentum_scan} shows that the magnitude of the ring observable increases with the jet velocity.  
As the momentum of the deposit energy-momentum current increases, it generates stronger velocity gradients in the fluid, producing higher vorticity values and leading to a stronger polarization along the vorticity ring. 

\subsection{ \label{sec:Results-Position-Centrality-Scan} The vortex ring's dependence on the initial position of the deposited energy-momentum current} 

\begin{figure*}[]
\includegraphics[scale = 0.32]{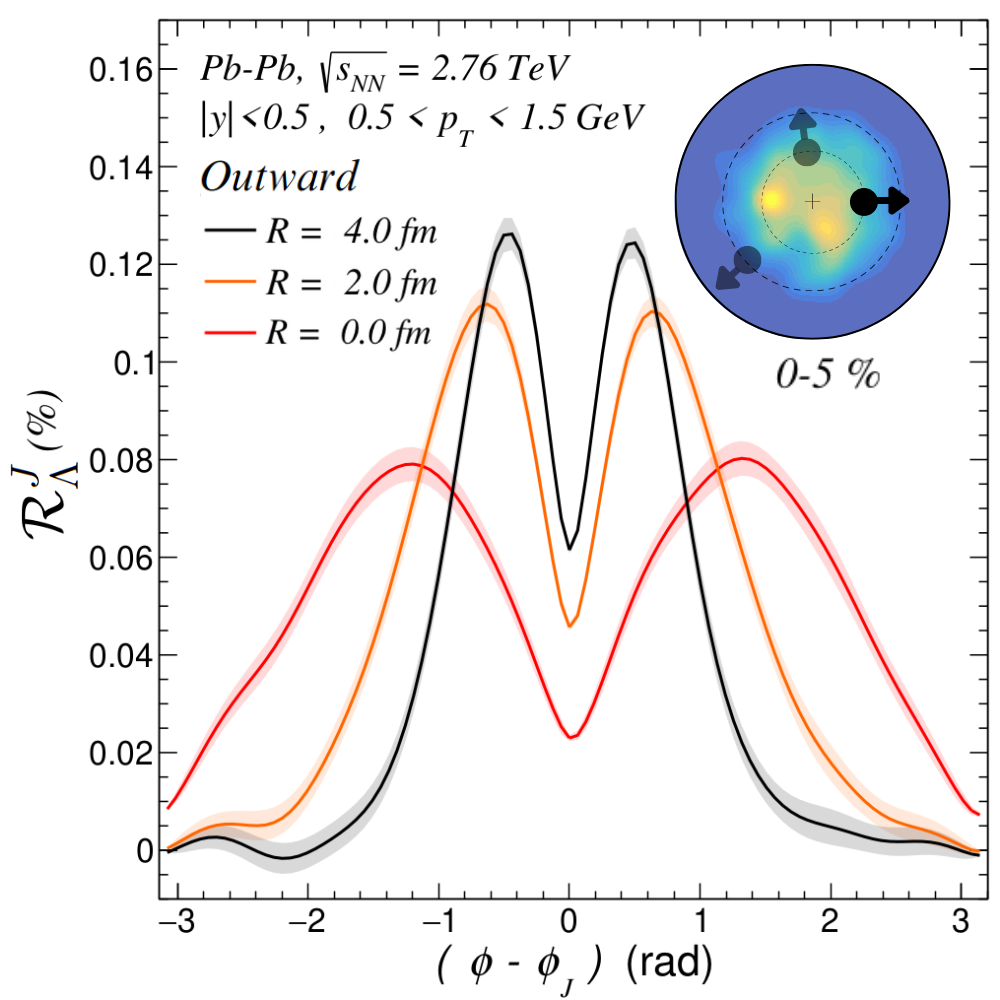}
\quad
\includegraphics[scale = 0.32]{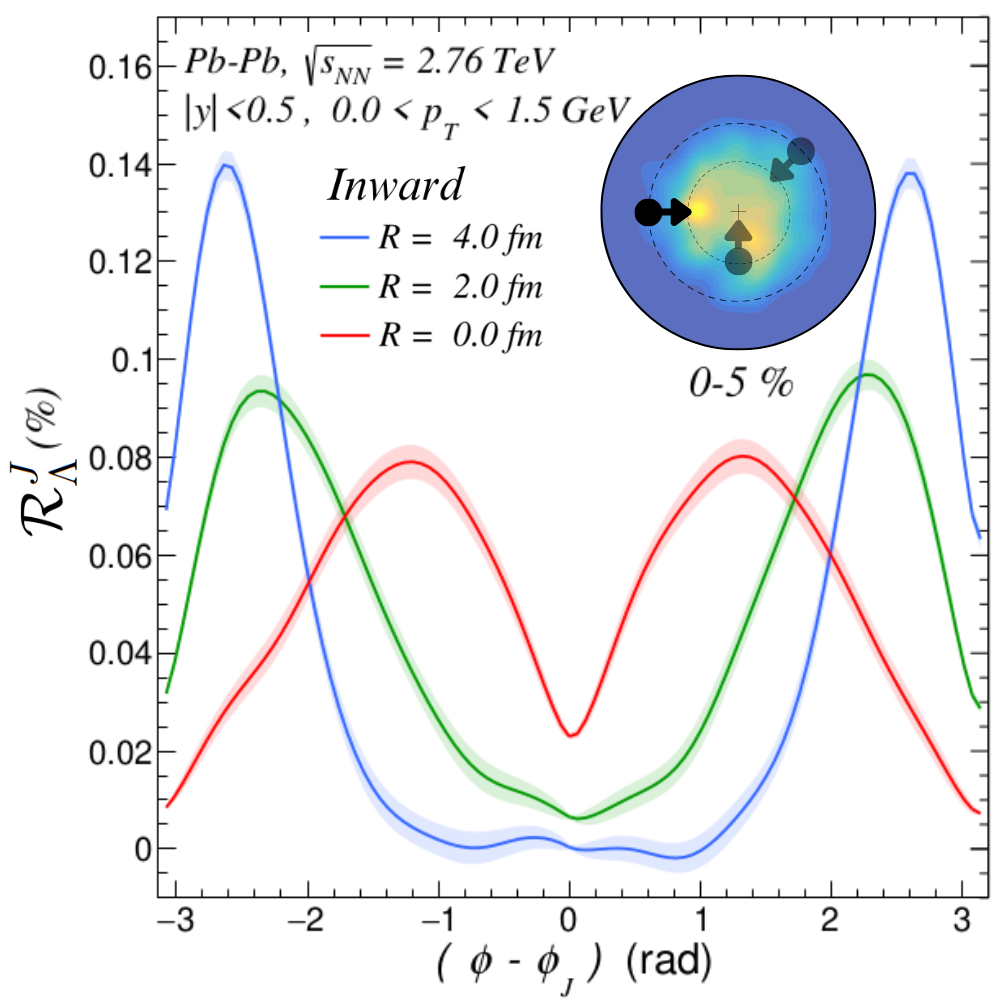}
\caption{\label{fig:position_scan_0_5} $\mathcal{R}^{J}_{\Lambda}$ distribution for different positions in the 0-5\% centrality class. In both panels, the thermalized jet is inserted in the medium in a position defined by the radius $R = 0, ~2 ~\text{and}~4$ fm. The direction of the jet $\phi_{J}$ was chosen at random and also completed the coordinate position by projecting the radius $R$ according to this direction. In the left panel, the direction of the energy-momentum currents deposited from the thermalized jet points outward to the center of the event. In contrast, in the right panel, the thermalized jet points towards to the center. The illustration inside the plots represents schematically the scenario studied, respectively, in which we applied fluctuating initial conditions and varied the position of insertion of the thermalized jet with a randomly selected momentum direction.}
\end{figure*}

\begin{figure*}[]
\includegraphics[scale = 0.32]{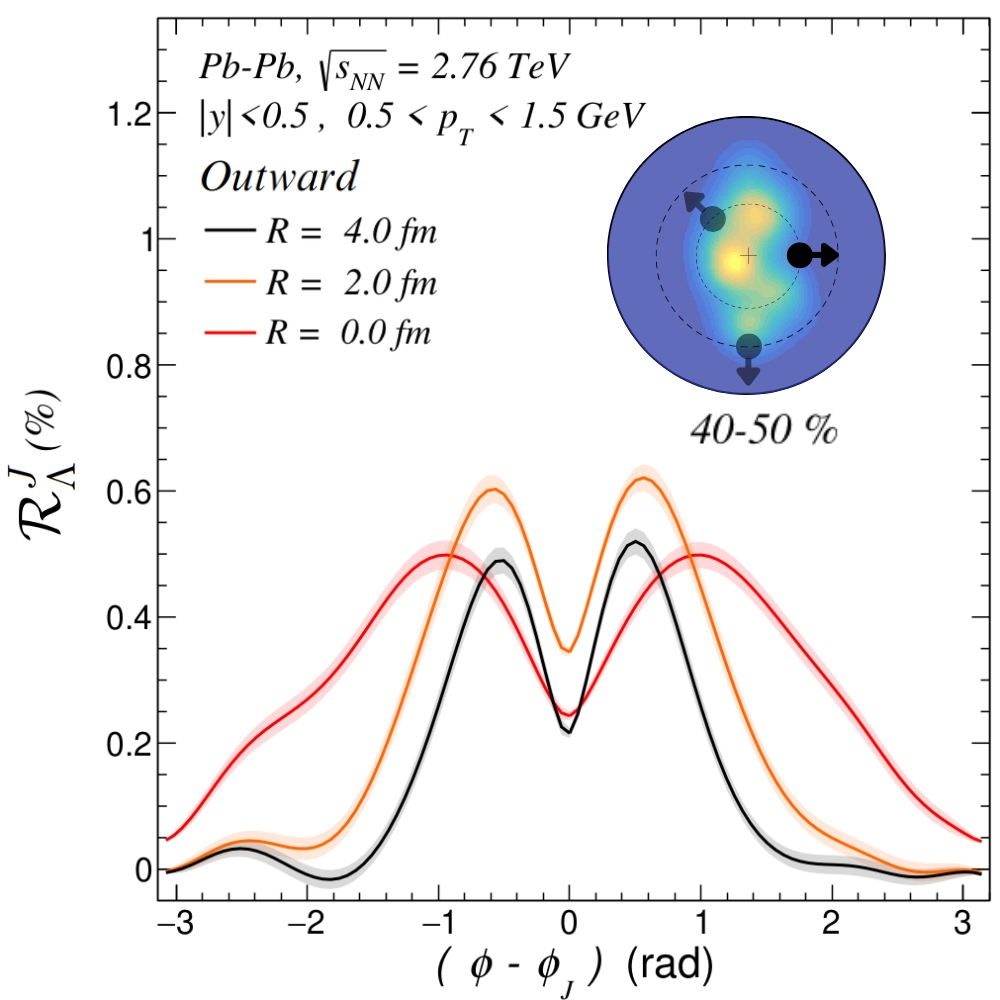}
\quad
\includegraphics[scale = 0.32]{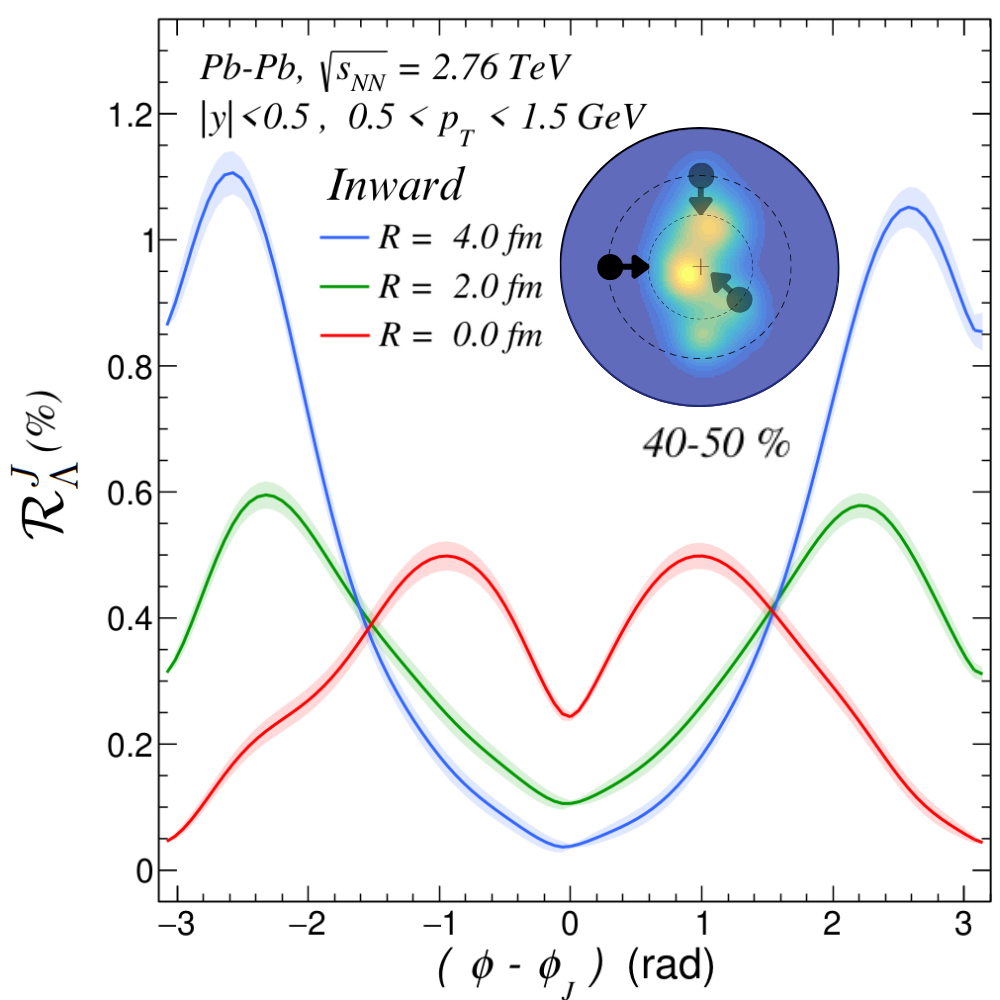}
\caption{\label{fig:position_scan_40_50} $\mathcal{R}^{J}_{\Lambda}$ distribution for different positions in the 40-50\% centrality class. In both panels, the thermalized jet is inserted in the medium in a position defined by the radius $R = 0, ~2 ~\text{and}~4$ fm. The direction of the jet $\phi_{J}$ was chosen at random and also completed the coordinate position by projecting the radius $R$ according to this direction. In the left panel, the direction of the energy-momentum currents deposited from the thermalized jet points outward to the center of the event. In contrast, in the right panel, the thermalized jet points towards to the center. The illustration inside the plots represents schematically the scenario studied, respectively, in which we applied fluctuating initial conditions and varied the position of insertion of the thermalized jet with a randomly selected momentum direction.}
\end{figure*}

Due to the fluctuations in realistic initial states, the media formed in each different collision is significantly different in the local temperature and flow velocity, which can directly affect the dynamics of the vortex ring created in that scenario. Based on this, in this section, we study how different initial positions for the deposited energy-momentum current would affect the azimuthal dependence of the ring observable in central (0-5\%) and semi-peripheral (40-50\%) Pb+Pb collisions. In this case, in order to account for the fluctuations in the IC profile, we performed 250 event-by-event (3+1)D hydrodynamic simulations with one energy-momentum deposition per event for both centrality bins.
Figures~\ref{fig:position_scan_0_5} and~\ref{fig:position_scan_40_50} show the ring observable as a function of the relative azimuthal angle to the jet direction ($\phi - \phi_{J}$). The ring observables are averaged over different insertion positions for the deposited energy-momentum currents. The propagation direction of the thermalized jets was chosen randomly from an interval of $[0, 2\pi)$ in each event.
In order to explore the different kinematic conditions characterized by the configurations formed between the direction of propagation of the quenched jet and the medium's flow, we define a position $(x,y)$ to insert the energy-momentum of the jet according to polar coordinates. Considering this picture, we used $R = 0, ~2,~\text{and} ~4$ fm to indicate the distance of the point where the energy-momentum currents were deposited from the center. Then, to complete the polar coordinate, we used the angles $\phi_J$ of the jet's momentum to project the $R$ distance into the horizontal and vertical axis. That description corresponds directly to the results expressed in the left panels of Figs.~\ref{fig:position_scan_0_5} and \ref{fig:position_scan_40_50}.
In that case, the direction of the jet velocity aligns with the system's expansion, and the propagation of the energy-momentum currents from the thermalized jet makes the vorticity ring evolve outward from the center of the event. On the other hand, in the right panels, we maintain the insertion position from the previous case but reverse the direction of the jet. Consequently, the propagation of the deposited energy-momentum currents opposes the system expansion at insertion time and leads to a vorticity ring that evolves towards to the center.

The results obtained in Figs.~\ref{fig:position_scan_0_5} and \ref{fig:position_scan_40_50} show two noteworthy features as follows. First, the maximum values of $\mathcal{R}^{J}_{\Lambda}$ increase with the $R$ distances of the energy-momentum deposition. Secondly, the distance between the peak angle, which characterizes the size of the ring, shows a stronger dependence on the positions of the jet energy deposition.
When the jet is inserted with a large $R$ and points outward to the center, the perturbation is already closer to the freeze-out surface. The velocity gradients induced by the jet are the largest at their insertion times and decrease with the hydrodynamic evolution.
The size of the ring, characterized by the angular distance between the two maxima in the ring observable, is smaller for a shorter evolution time because of the shorter time available for the vorticity ring to expand in the medium. Now, in the case where the jet insertion was done propagating toward the center, the dynamics that generate the observed signal are similar to the previous case. However, due to the fact that the jet energy travels in a direction that opposes the system expansion, the induced velocity gradients are much larger. They will dominate the signal contribution despite the longer evolution time. For this case, the velocity gradients will reach the freeze-out surface with a greater contribution in a region behind the jet, which can be seen in the peak angles that are close to $(\phi-\phi_J)=\pi$ or $-\pi$.
In addition to that interpretation, the left panel of Fig.~\ref{fig:position_scan_40_50} shows a different behavior in comparing the magnitudes of the results obtained for $R=4$ fm and $R=2$ fm. For that comparison, it is important to consider the amount of matter that will hydrodynamically interact to form the vorticity ring. As in peripheral events when the jet is inserted in a position that is very close to the boundaries of the system, it is expected that the interaction of the thermalized jet with that region of the fluid to be small, which consequently affects the formation of the vorticity ring and decreases the vorticity field created.

\begin{figure*}[]
\includegraphics[scale = 0.32]{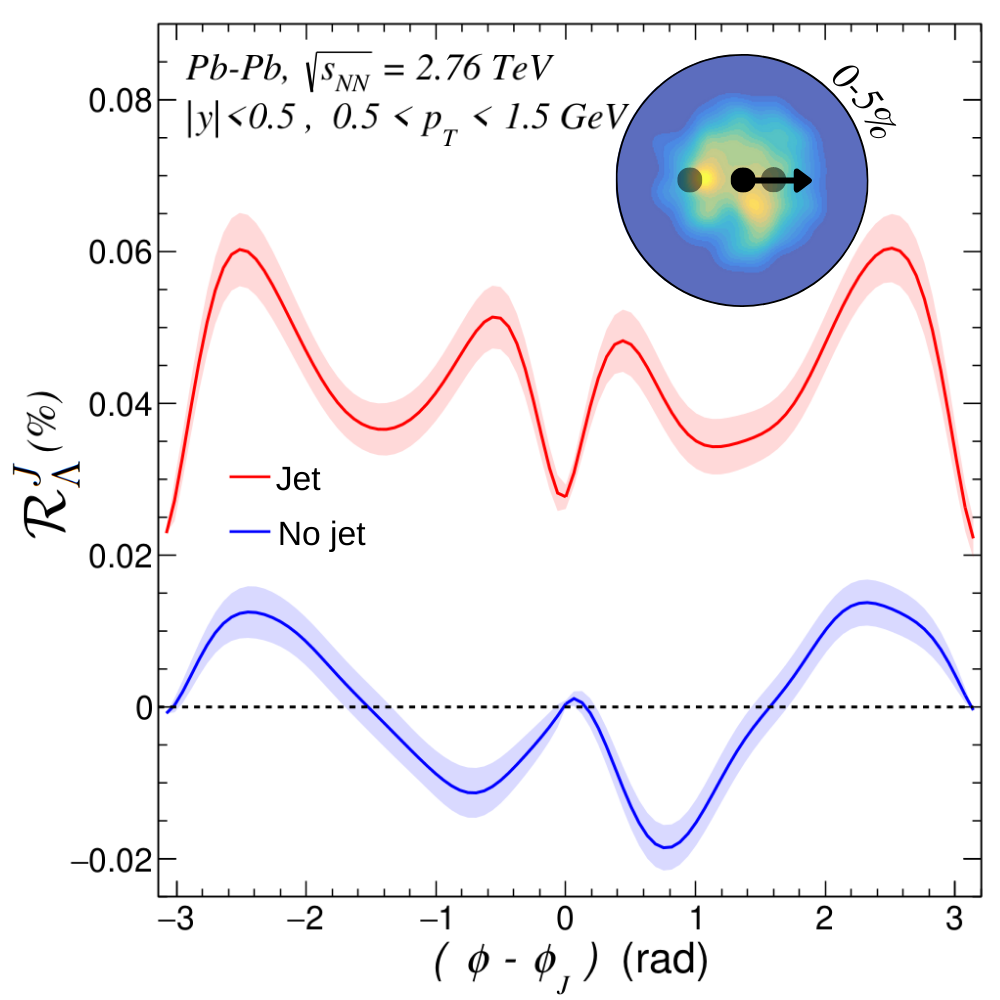}
\quad
\includegraphics[scale = 0.32]{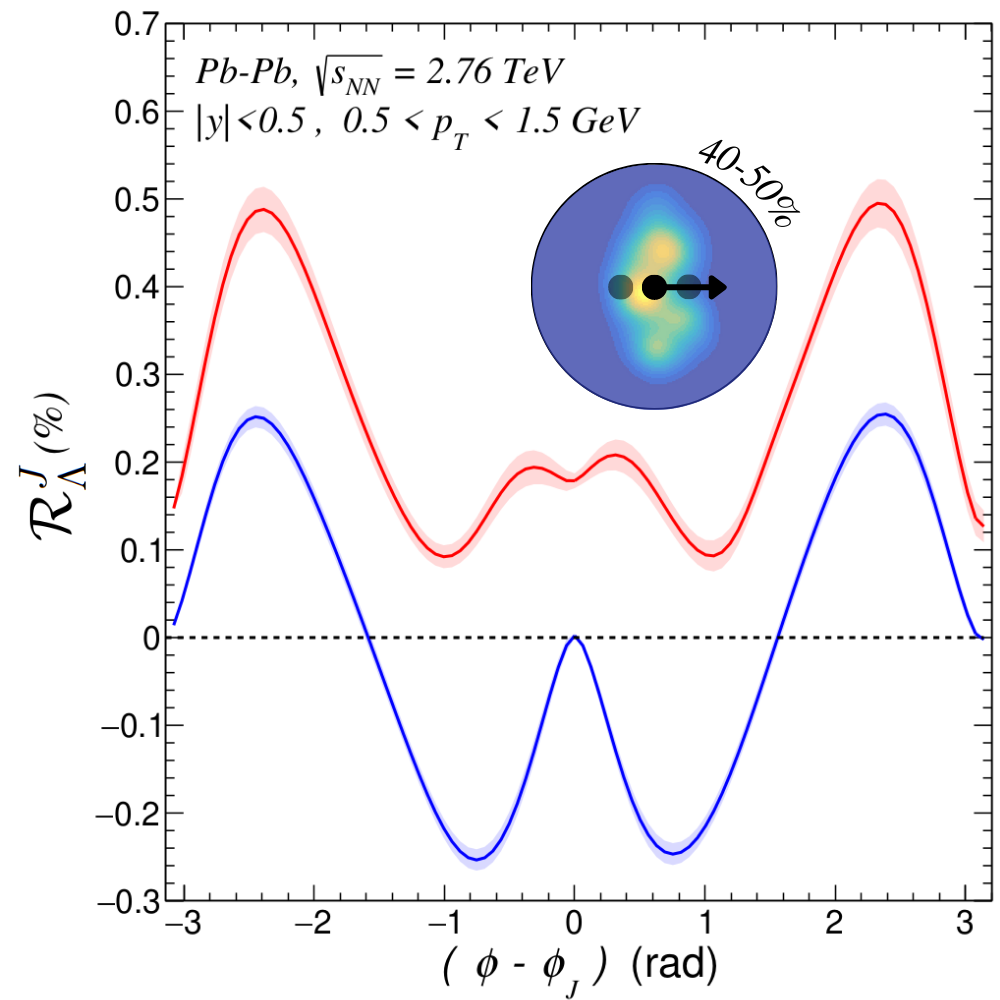}
\caption{\label{fig:sorted_x} $\mathcal{R}^{J}_{\Lambda}$ as a function of the azimuthal angle relative to the jet's direction for two different centralities: (left) $0-5\%$ and (right) $40-50\%$. For these results, the position where the jet is inserted was chosen randomly according to the normalized entropy distribution of the central slice projected into the $x$-axis ($y = 0 $ fm and $\eta = 0$) of each initial condition profile. The illustration inside the plots represents schematically the scenario studied, in which we applied fluctuating initial conditions of each respective centrality class and varied the position of insertion of the thermalized jet with the momentum always pointing positive along the x-axis.}
\end{figure*}

Considering the position dependence observed up to this point, in order to make a more realistic prediction for jet-induced vortex ring in heavy-ion collisions, 
it is important to account that the production position of high-energy jets in real collisions fluctuates from one event to another according to the binary collision profile. 
A complete analysis of that scenario should account for jets being produced in any part of the medium and propagating with a direction that should vary in each event. In that context, motivated by results that will be presented in the next section, it is  expected that the main contribution of the signal should be caused by events where the momentum of the thermalized jet is aligned with the medium's flow. Additionally to that, due to the computational limitation of performing such an analysis for all directions, we prioritized a study in the $x$-axis.
For this, we varied the positions of the energy-momentum deposition from the quenched jets, considering a probability distribution in the $x$-axis given by the normalized energy density profile of the initial condition, which serves as a proxy for the distribution of binary collisions.
For these calculations, we consider an energy profile along the $x$ direction ($y = 0$ and $\eta_{S} = 0$) for each fluctuating initial condition and insert the thermalized jet into the hydrodynamic medium with a momentum oriented along the $\hat{x}$ direction. 
Fig.~\ref{fig:sorted_x} shows the ring observable as a function of the relative azimuthal angle $(\phi-\phi_{J})$ for 0-5\% (left) and 40-50\% (right) Pb+Pb collisions at 2.76 TeV. The red curves correspond to the ring observable obtained with the jet insertion, and the blue curves to the background results, \textit{i.e.} without the jet's energy-momentum deposition. For the results obtained from events with energy-momentum currents from thermalized jets, the size of the vorticity ring varies in each event due to the variation in the position of the jet insertion. Because of this, the pattern of a characteristic vortex ring, which is presented in all the previous analyses with azimuthal dependence, is no longer visible. For all these previous analyses, the vortex ring was generated under the same conditions, respective to each study, and then had approximately the same size. However, for the study performed in Fig.~\ref{fig:sorted_x}, the different locations under which each vorticity ring was created led to very different vortex rings. The final signal is an average over those different contributions. Additionally, for the results where no energy-momentum currents were inserted, the signal of the background is dominated by the anisotropic expansion contribution. The larger magnitude observed in peripheral collisions can be attributed to the more intense anisotropy generated during these collisions and to the shorter lifetime of the fireball.
The main message of this analysis is that, even with a more realist scenario considering the binary collision distribution and also a non-vanishing influence of the transverse expansion, the final signal calculated through $\mathcal{R}^{J}_{\Lambda}$ is sensitive to the effects of the vortex rings created and can be easily differentiated from the signal of events with no jet.

\begin{figure}[]
\includegraphics[scale = 0.32]{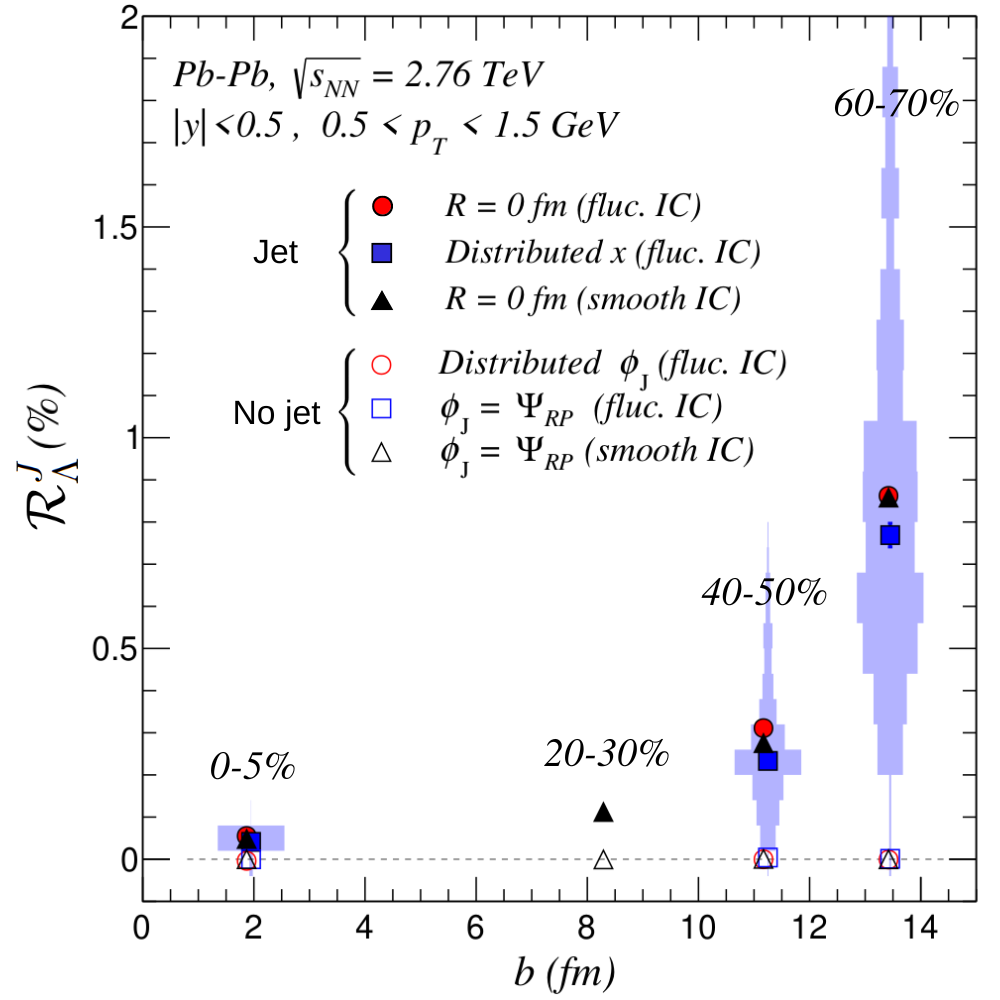}
\caption{\label{fig:Rp_vs_b} Values of integrated $\mathcal{R}^{J}_{\Lambda}$ as a function of the impact parameter $b$ for different scenarios, including those with and without the insertion of the thermalized jet. The results were obtained for different conditions of simulation: jet inserted in the center $R=0$ with fluctuating IC; varying the jet insertion position in the x-axis in fluctuating IC; and jet in the center with smooth IC. The shaded area corresponds to the distribution of the event-by-event $\mathcal{R}^{J}_{\Lambda}$ values obtained from the 250 events of the \textit{Distributed x (fluc. IC)} analysis.}
\end{figure}

To complement and directly compare our previous results from Figs.~\ref{fig:smooth-vs-ebe}, \ref{fig:position_scan_0_5}, \ref{fig:position_scan_40_50} and \ref{fig:sorted_x}, we investigated the influence of the centrality class on the integrated value of $\mathcal{R}^{J}_{\Lambda}$.
Figure~\ref{fig:Rp_vs_b} presents the azimuthal-integrated ring observable as a function of the impact parameter $b$ for different scenarios of energy-momentum deposition from the quenched jets.
The red and blue points in Fig.~\ref{fig:Rp_vs_b} represent the mean of the results obtained from the event-by-event analysis, and the error bar corresponds to the statistical error of the values obtained. 
The points in the same centrality bin are slightly offset from each other in $b$ for clear visualization. 
The first case, labeled as \textit{$R=0$ fm (fluc. IC)}, represents the azimuthal-integrated signal of the $\mathcal{R}^{J}_{\Lambda}$ in the scenario $R=0$ studied in Fig.~\ref{fig:position_scan_0_5} and Fig.~\ref{fig:position_scan_40_50}. The second case, \textit{Distributed x (fluc. IC)}, corresponds to the conditions applied in Fig.~\ref{fig:sorted_x}. In this case, a shaded blue area is also presented and represents the distribution of the integrated values for the 250 fluctuating events performed. The last case, denoted as \textit{$R=0$ fm (smooth IC)}, represents the analysis performed with a single smooth initial condition, in which the jet energy-momentum was inserted at a position defined by $x=y=0$, with momentum pointing to the $+x$ axis. Results obtained for each analysis scenario in the absence of jet insertion are also presented and plotted with unfilled markers. 
For these cases, the legend provides the corresponding reference angles in the calculation for $\mathcal{R}^{J}_{\Lambda}$.

The results presented in Fig.~\ref{fig:Rp_vs_b} show that the magnitude of the ring observable induced by the jet-medium interactions increases in peripheral collisions. This dependence can be assigned to a combination of factors. Firstly, the ratio between the deposited energy and background energy density in peripheral collisions is larger than in central collisions, leading to bigger velocity gradients in the peripheral events. Additionally, peripheral collisions have a shorter total evolution time, resulting in an earlier particlization when the vorticity field is stronger. The azimuthal-integrated ring observable is consistent with zero for the results of no energy-momentum currents deposition from quenched jets. Therefore, the proposed ring observable has a null background, and the jet-medium excitation would induce any non-zero measurements. The general picture provided by our analysis demonstrated that the ring observable is shown to be robust not only with respect to different insertion positions for the jet but also when comparing event-by-event analysis with a smooth initial condition analysis.

\subsection{The ring observable's sensitivity on jet alignment} 

A final remaining variable that needs to be explored to represent a real scenario accurately is the addition of a fluctuating direction of the quenched jet on top of the position fluctuations. 
This is important because, in all results presented up to this point, we assumed that the direction of the thermalized jet and the direction of the flow of the system were always parallel. 
Even in the scenario when the energy-momentum current is against the radial flow, analyzed in the previous subsection and shown in Figs. \ref{fig:position_scan_0_5} (right) and \ref{fig:position_scan_40_50} (right), the axis that defines the propagation of the jet and the flow are the same, but pointing in opposite directions. However, in reality, it is possible for the jet to be oriented differently with respect to the flow axis. To investigate the effects of misalignment between the jet energy-momentum current and underlying flow velocity, we perform event-by-event calculations and fix the insertion position of the energy-momentum current of the jet at $x=2$ fm and $y=0$. We vary the jet's azimuthal angle ($\phi_{J}$) 
in each of 180 fluctuating events by an interval of $0<\phi_{J}<2\pi$ with a space of $\Delta \phi = 2\pi/180$, thereby breaking the alignment between the jet direction and flow axis.

\begin{figure*}[]
\centering
\includegraphics[scale = 0.2]{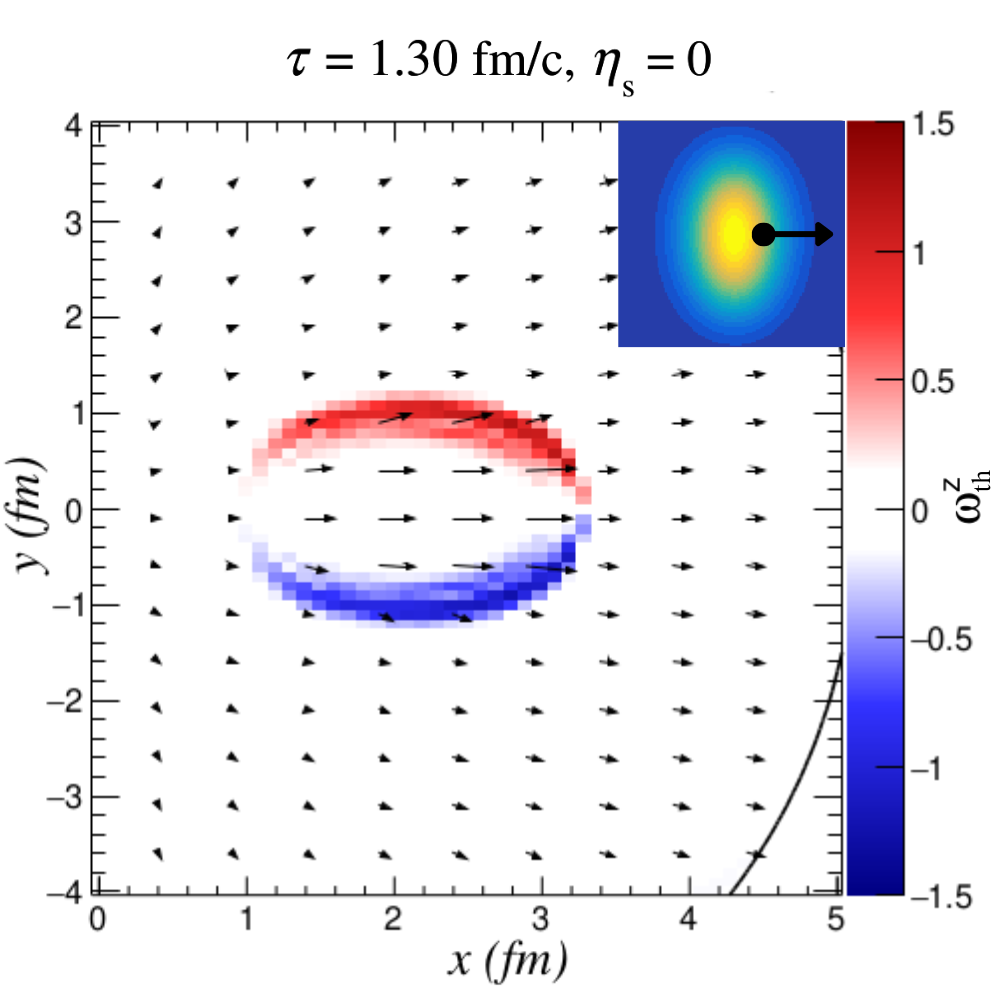}
\quad
\includegraphics[scale = 0.2]{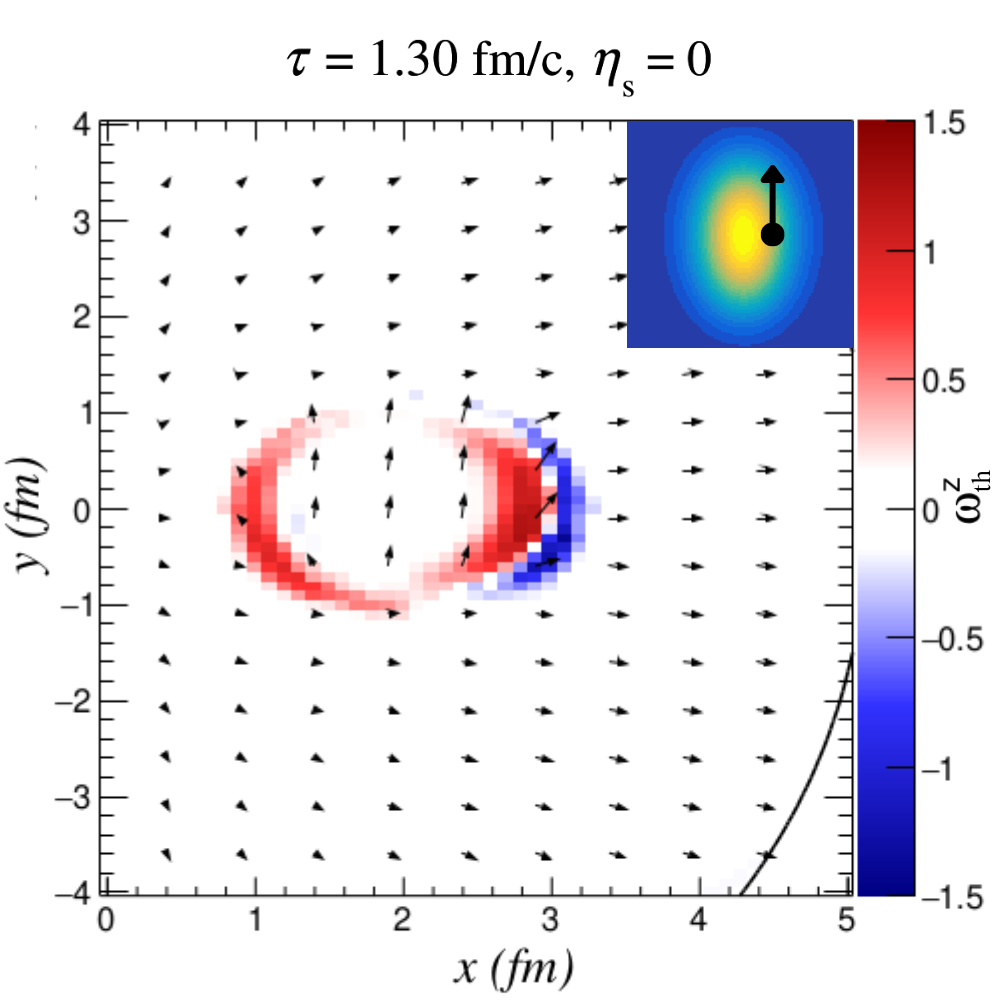}
\quad
\includegraphics[scale = 0.2]{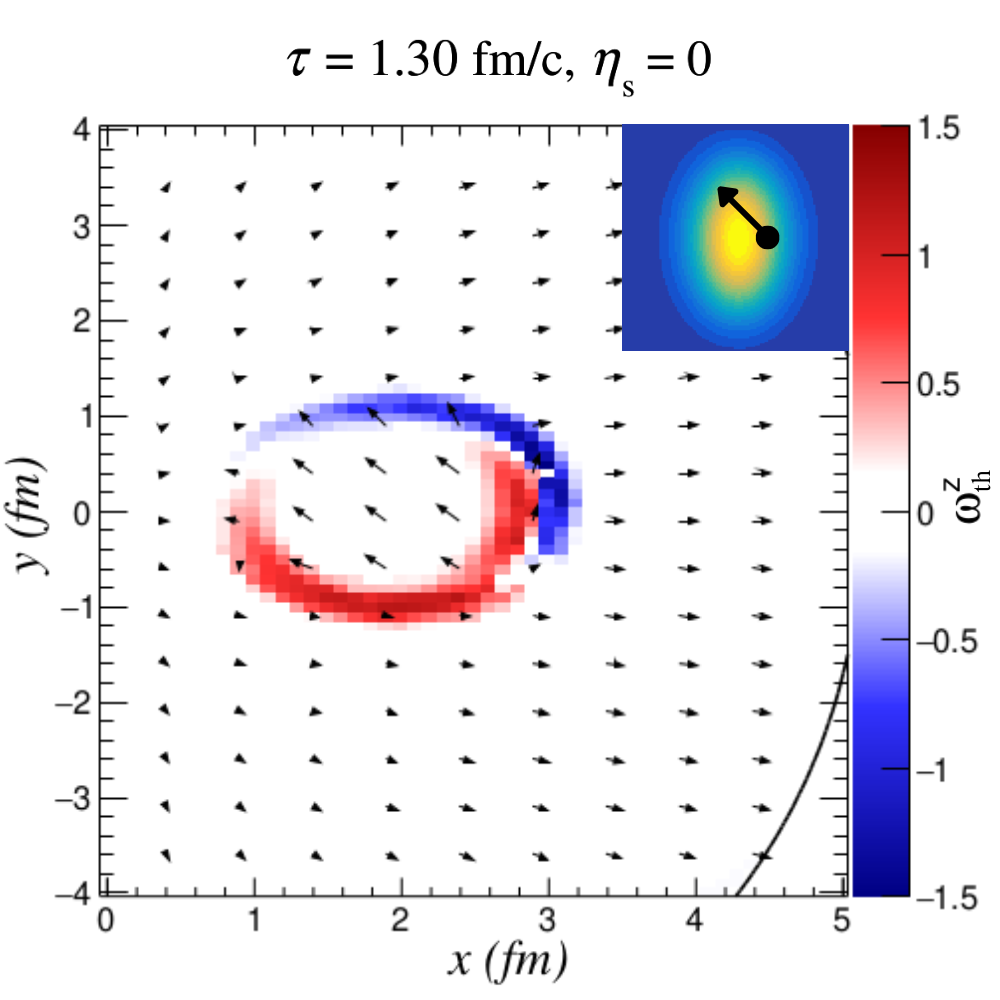}
\qquad
\includegraphics[scale = 0.2]{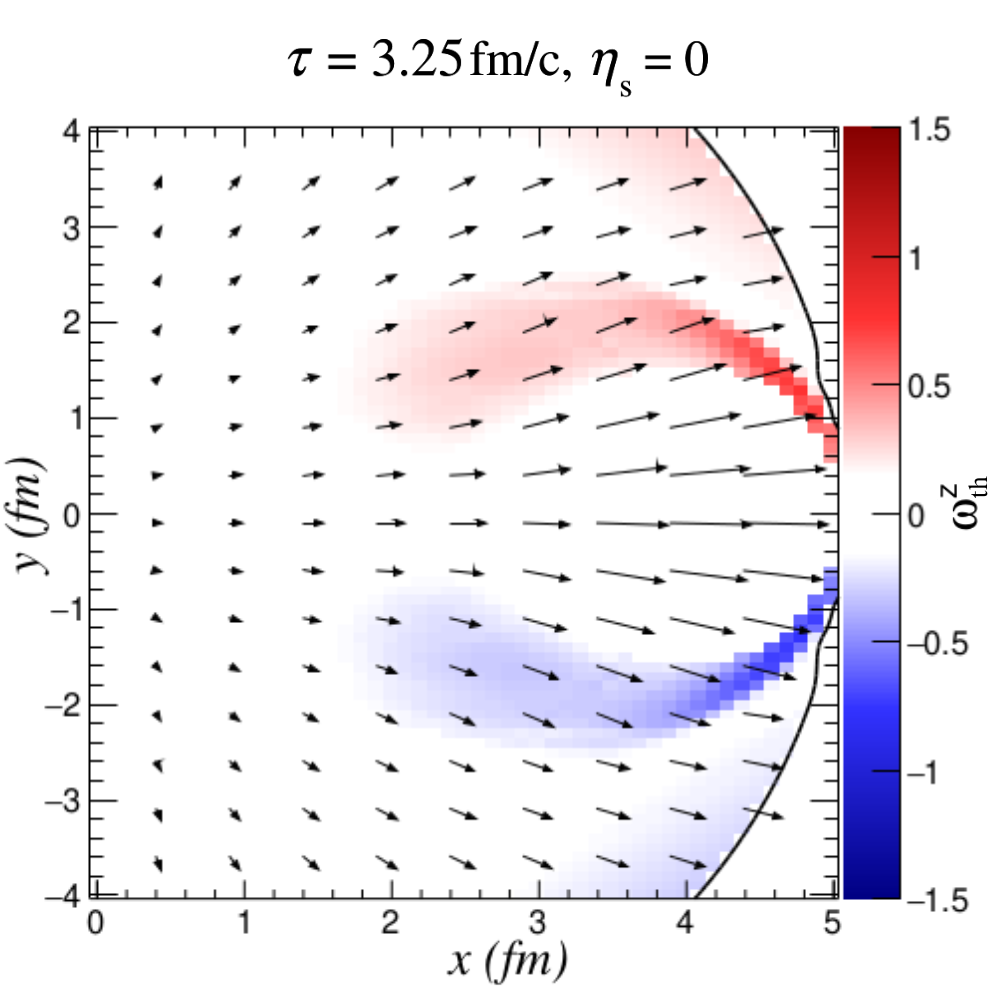}
\quad
\includegraphics[scale = 0.2]{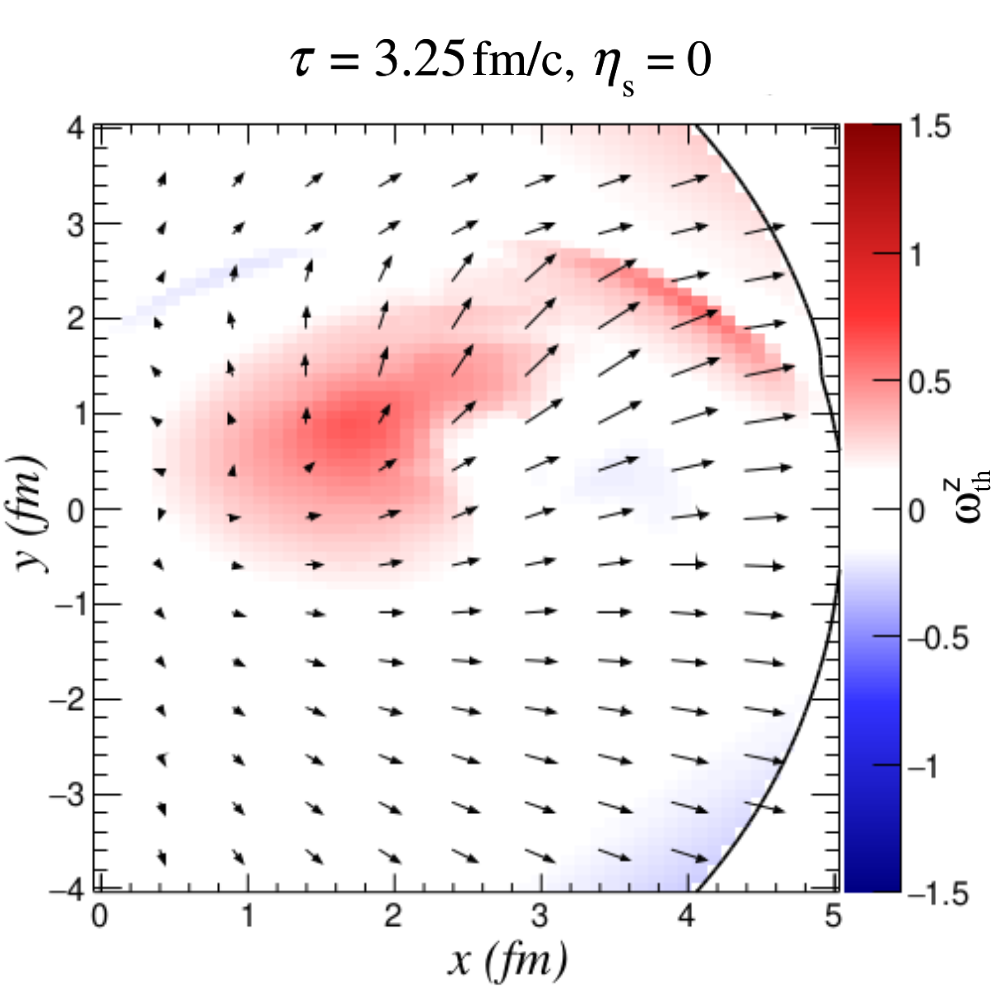}
\quad
\includegraphics[scale = 0.2]{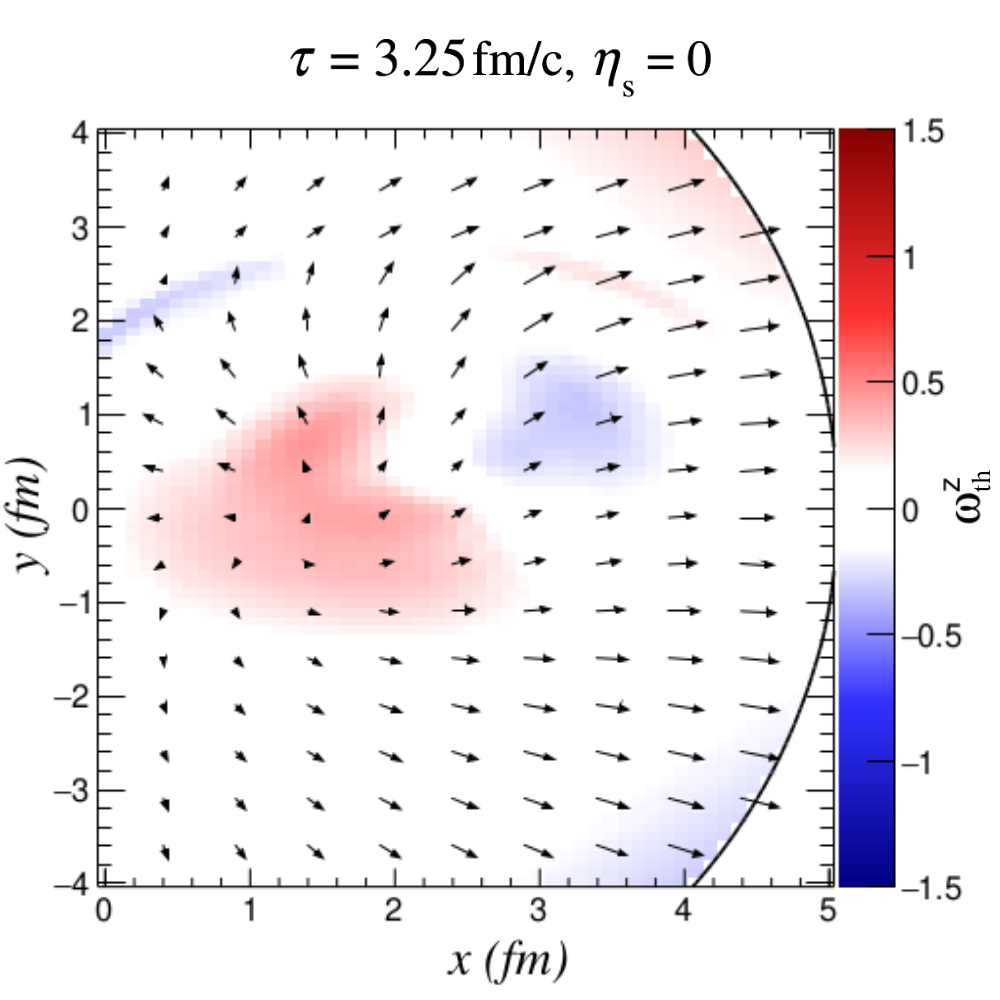}
\caption{\label{fig:jet-aligment} Vortex ring formed by the energy-momentum currents of the thermalized jet for time steps $\tau = 1.30 ~\text{fm}/c$ and $\tau = 3.25 ~\text{fm}/c$ of the hydrodynamic evolution. The panels show the spatial distributions of the $z$-component of the thermal vorticity, $\omega^z_\mathrm{th} = \epsilon^{z \rho\mu\nu} u_\rho \varpi_{\mu \nu}$, at mid-rapidity. The arrows indicate the fluid's traverse velocity. For all the results, the energy-momentum currents were inserted at $y=0$ fm and $x=2$ fm in a smooth initial condition in 40-50\% centrality class. In the upper panels, there is an insert scheme to represent the orientation of the jet inside the event. The left panel represents the evolution of the vorticity generated by a thermalized jet with momentum pointing positive in the horizontal axis ($\phi_J = 0$ rad). The central panels represent the evolution of the vorticity generated by a thermalized jet with momentum pointing positive in the vertical axis ($\phi_J = \pi/2$ rad). The right panel represents the evolution of the vorticity generated by a thermalized jet with momentum oriented by an angle of $\phi_J = 3\pi/4$ rad relative to $+\hat{x}$ direction. The black solid line present in the plots corresponds to the freeze-out surface.}
\end{figure*}

The azimuthal-integrated ring observable obtained for 40-50\% Pb+Pb collisions yield a value of $\mathcal{R}^{J}_{\Lambda} = 0.1282 \pm 0.0110 \: (\%)$.
This resulting signal corresponds to approximately 41\% of the signal compared to the scenario when the jet is inserted at the system's center, and the jet's momentum is aligned with the flow velocity. Such a decrease can be assigned to the deformation of the toroidal structure of the vortex ring caused by the different configurations between the ring's expansion direction and the medium's flow direction. To illustrate this outcome, Figure~\ref{fig:jet-aligment} presents some of those configurations. As a consequence of the deformation, the ring pattern of vorticity that induces the polarization of the hyperons is destroyed. It becomes clear by looking at a later evolution time shown in the lower panels of Fig.~\ref{fig:jet-aligment}. While the left panel shows a well-defined pattern for the vorticity created in the thermalization of the energy-momentum currents of the jet, the center and right panels show a highly deformed pattern for the vorticity.  However, despite the fact that the signal decreases significantly, it is worth emphasizing that it is still nonzero and has a measurable magnitude for $\mathcal{R}^{J}_{\Lambda}$.

\subsection{ \label{sec:Results-expansion} Collective expansion effects on the ring observable}

Lastly, we shift our focus to perform an exploratory study on the vortex ring pattern induced by the expansion of the anisotropic hydrodynamic flow. Regarding the topic of medium expansion, it is possible to evaluate the evolution of the quark-gluon plasma in two distinct scenarios: longitudinal and transversal expansion. On the one hand, the longitudinal expansion accounts for the dynamics that can be observed along the $\eta_s$ direction. In contrast, the transverse dynamics is related to the expansion in the transverse $x-y$ plane. In both scenarios, a local non-zero vorticity is generated for the cases in which its dynamics present an anisotropic nature. The resulting vorticity will then induce the polarization of the $\Lambda$ hyperons emitted from the event. The longitudinal expansion will be responsible for inducing the polarization in the directions perpendicular to the longitudinal axis, which means $\hat{y}$ and $\hat{x}$ directions, and the transversal expansion will be responsible for inducing the polarization in the longitudinal direction. In this section, to focus only on the effects of medium expansion, we excluded the contributions that could appear from the energy-momentum currents of quenched jets and the global angular momentum generated in the collision. The only contribution to polarization will be the vorticity generated in the anisotropic expansion of the medium.

For this study, we used a smooth initial condition equivalent to 0-5\% centrality class, which was hydrodynamically evolved without any jet insertion. We also account for recent works that indicate an additional contribution to spin polarization due to a shear-induced term \cite{Becattini_2021, Liu_2021}. For both works, the mean spin vector of the $\Lambda$ hyperons is defined with a similar relation compared to Eq. \eqref{eq:mean_spin} 

\begin{equation}\label{eq:spin}
    S^{\mu}(p) = - \dfrac{1}{4m} \dfrac{\int_{\Sigma} \text{d}\Sigma \cdot p \: n_{F}(1-n_{F}) \mathcal{A}^{\mu}}{\int_{\Sigma} \text{d}\Sigma \cdot p \: n_{F}},
\end{equation}
where $\mathcal{A}^{\mu}$ corresponds to the different derivations for the shear contribution achieved in each analysis \cite{Becattini_2021, Liu_2021}.

Phenomenological studies already start to evaluate the difference between both definitions \cite{Yi_2021, Alzhrani_2022}, which represents an important step in understanding the vorticity-spin coupling for relativistic hydrodynamic systems. Following this trend, the definitions of spin that we used to calculate the final polarization of the $\Lambda$ hyperons in this analysis are represented below. 

In Ref. \cite{Becattini_2021}, the complementing term of Eq. \eqref{eq:spin} is given by

\begin{equation}\label{eq:BBP}
    \mathcal{A}^{\mu}_{BBP} = \epsilon^{\mu \nu \sigma \tau}  \left( \dfrac{1}{2} \varpi_{\nu\sigma}p_{\tau} + \dfrac{1}{E}\hat{t}_{\nu} \xi_{\sigma\lambda} p^{\lambda}p_{\tau} \right),
\end{equation}
where $\hat{t} = (1, 0, 0, 0)$ corresponds to a global vector, $\varepsilon$ represents the energy density and $\xi_{\sigma \lambda}$ is the thermal shear tensor defined as
\begin{equation}
    \xi_{\mu\nu} = \dfrac{1}{2}(\partial_{\mu}\beta_{\nu} + \partial_{\mu}\beta_{\nu})
\end{equation}
with $\beta_\mu$ corresponding to the temperature four-vector $\beta_\mu = u_\mu/T$.

On the other hand, in Ref. \cite{Liu_2021,Yi_2021}, the definition of the mean spin vector is complemented by the term
\begin{equation}\label{eq:LY}
    \mathcal{A}^{\mu}_{LY} = \epsilon^{\mu \nu \sigma \tau}  
    \left[ \dfrac{1}{2} \varpi_{\nu\sigma}p_{\tau} + \dfrac{1}{E}u_{\nu} \xi_{\sigma\lambda} p^{\lambda}_{\perp}p_{\tau} \right],
\end{equation}
where $p_{\perp}^{\lambda}$ corresponds to an additional transverse projection operator defined as

\begin{equation}
    p_{\perp}^{\lambda} = p^{\lambda} - (u \cdot p) u^{\lambda}.
\end{equation}

\begin{figure}[]
    \includegraphics[scale = 0.26]{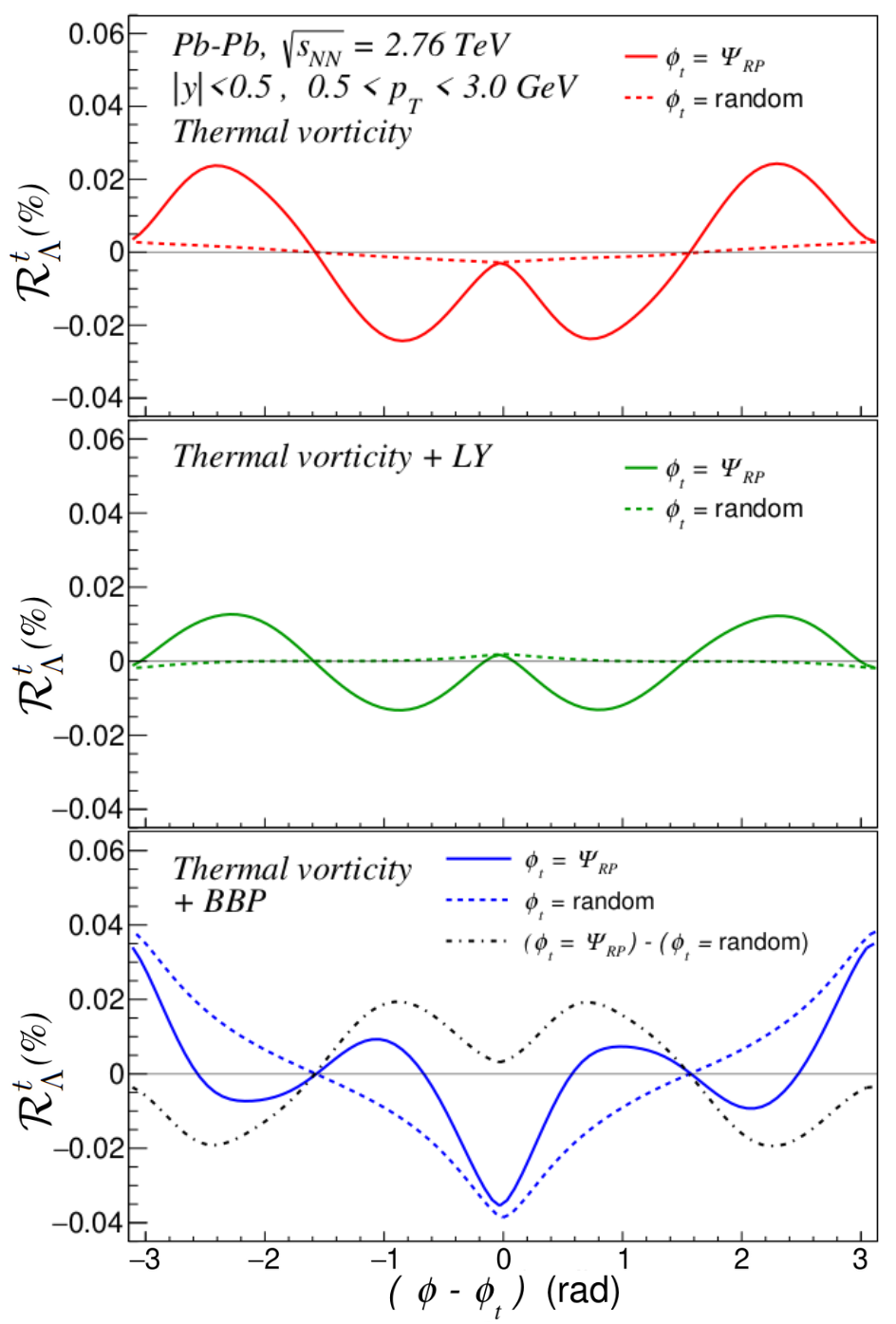}    
    \caption{ \label{fig:expansion} Values of $\mathcal{R}^{t}_{\Lambda}$ with different definitions of the polarization induced by the expansion of the system for 0-5\% Pb+Pb collisions. These results correspond to an analysis where a smooth initial condition was used. The solid lines represent $\mathcal{R}^{t}_{\Lambda}$ calculated for the trigger angle $\phi_{t}$ in the reaction plane $\Psi_{RP}$ and the dashed lines correspond to results obtained from an analysis where $\mathcal{R}^{t}_{\Lambda}$ was averaged over 1000 different $\phi_{t}$ sampled from a uniform distribution defined by the interval $[0, 2\pi)$. The top panel corresponds to results where the polarization is induced by the local thermal vorticity. The middle panel corresponds to results where the polarization is defined by the sum of the thermal vorticity contribution with the shear-induced term LY from \cite{Liu_2021}. The bottom panel corresponds to results where the polarization is defined by the sum of the thermal vorticity contribution with the shear-induced term BBP from \cite{Becattini_2021}. The black dot-dashed line in the bottom panel represents a proxy signal that only displays contributions of the transverse expansion.}
    
\end{figure}

We compared the azimuthal distribution of the ring observable with and without the shear-induced polarization and also for both definitions that included a shear term. The comparison is shown in Fig.~\ref{fig:expansion}, with a $p_{T}$ interval in the range of $0.5 < p_{T} < 3.0$ GeV$/c$, which prioritizes the signal of the background as could be seen in the left panel of Fig.~\ref{fig:sz-no-jet}. 
At the beginning of this work, in \ref{sec:Methodology}, we showed that summing the signals calculated with different $\phi_{t}$ would average zero and vanish with the transverse contribution (see the red dashed lines on Fig.~\ref{fig:smooth-vs-ebe}).
Since this method is based on azimuthal correlations in the $x-y$ plane, it should not affect the distribution of polarization in the longitudinal plane. The result of this statement is that the contribution of polarization induced by the longitudinal expansion will not average zero as the transverse contribution does. Such an observation, characterized through the calculus of the ring observable, presents a characteristic that could be used in order to decouple the longitudinal contribution from the transverse one. 
This process is used in our analysis by averaging $\mathcal{R}^{t}_{\Lambda}$ over 1000 calculations with $\phi_{t}$ being selected at random from a uniform distribution of the interval $[0, 2\pi)$. The resulting signal obtained is shown with $\phi_{t} = \text{random}$ in Fig.~\ref{fig:expansion}.

For the considerations of our results, expressed in Fig.~\ref{fig:expansion}, the signal obtained with $\phi_t = \Psi_{RP}$, accounts for both transversal and longitudinal expansions, while for $\phi_t = \text{random}$ the transversal contribution is averaged out and the signal is attributed just to the longitudinal contributions. By looking at Fig.~\ref{fig:expansion} it is possible to see the results obtained for each definition of polarization in a different panel. For the results expressed in the upper and central panels, it is possible to verify that signal of $\mathcal{R}^{t}_{\Lambda}$ attributed just to the longitudinal expansion (dashed lines) presents a small magnitude relative to the result where the transversal contribution also accounts (solid line). Such an observation shows that for the polarization induced by thermal vorticity only and the one defined by thermal vorticity plus the shear-induced term from Ref.~\cite{Liu_2021}, the contribution of transversal expansion is the dominant one. In contrast, a great difference is observed for the results where the polarization is defined by the inclusion of the shear-induced term derived in Ref.~\cite{Becattini_2021}. In the lower panel of Fig.~\ref{fig:expansion}, the comparison of the dashed line with the solid one reveals a remarkable increase in the longitudinal contribution. In order to compare the transversal and longitudinal contributions produced by that definition of polarization, we subtracted the signal obtained with $\phi_t = \text{random}$ from the signal where $\phi_t=\Psi_{RP}$. 
This method is used as a proxy to isolate the contribution that should be caused just by the transversal expansion. The result of that subtraction is expressed in the black dash-dotted line. Comparing the magnitudes of the signal induced by just the longitudinal dynamics (dashed line) with the proxy-signal attributed just to the transversal expansion (dash-dotted line), the maximum value achieved for the longitudinal contribution presents a magnitude that accounts for almost twice the maximum observed for the transverse signal. This observation indicates a dominance of the longitudinal contribution and represents an inversion of the hierarchy presented by the other two definitions of polarization. 
It is also important to highlight that the qualitative properties of the black dash-dotted line of the lower panel in Fig.~\ref{fig:expansion} are very similar to the proprieties of the signal represented by the solid lines in the top and center panels. All the results present the same azimuthal dependence, which is characteristic of a signal induced by the transverse anisotropic expansion of the medium. The only striking difference observed in the comparison is the sign presented by the curve in the lower panel, which is opposite to the sign observed for curves in the center and upper panels. That difference is in accordance with the properties observed in \cite{Becattini_2021_2}, which indicates an inversion of the polarization signal for the polarization calculated using the shear-induced contribution of Ref.~\cite{Becattini_2021}.

\section{Conclusions}

In this work, we systematically studied how the energy-momentum currents of quenched jets can induce a ring pattern of vorticity in the QGP medium. 

To expand on the previous work~\cite{Serenone2021}, we quantify the effects of fluctuating initial conditions on the ring observable of $\Lambda$ hyperons. We computed the polarization of the $\Lambda$ hyperons due to the vorticity generated by the thermalization of the quenched jet and applied this framework to evaluate different aspects of jet-medium interactions. 

Our systematic study clearly demonstrates that the ring observable is a highly sensitive tool to probe the dynamics of the QGP and jet-medium interactions. The results showed that $\mathcal{R}^{J}_{\Lambda}$ is sensitive to the momentum of the quenched jet, the specific shear viscosity of the medium, and also to the position and direction at which the energy-momentum current is deposited. These observations establish the ring observable as a powerful tool for studying the dynamics and interactions of the extreme matter created in high-energy heavy ion collisions. Furthermore, we showed that even with fluctuations in initial conditions, the effect calculated by the ring observable $\mathcal{R}^{J}_{\Lambda}$ has the same magnitude as reported by ALICE and STAR for the global $\Lambda$ polarization \cite{ALICE, STAR}. With this comparison, our calculations suggest that this observable can be measured, providing a clear signal of the thermalization of the energy lost by a quenched jet. 

We also showed that, for impact parameter $b \neq 0$ fm, there was a contribution of polarization caused by the anisotropic transverse expansion of the fluid, which should increase for peripheral events. We point out that this expansion effect significantly influences the signal attributed to the jet, which requires a method to separate the different contributions to visualize just the effects caused by the jet thermalization.
To accomplish this, we conducted a study that randomized the direction of the jet momentum and demonstrated the efficacy of this method in providing a decoupling of the jet signal from the $v_2$ signal. Additionally, we performed a complementary analysis to explore further the polarization induced by system expansion. Our results showed that the ring observable is a valuable tool for decoupling the polarization generated by transverse and longitudinal expansion, which also represented an experimental opportunity for testing the proposed shear-induced terms in the description of polarization.

\begin{acknowledgments}
VHR, DDC, JT and GT are supported by FAPESP projects 17/05685-2 (all), 21/10750-3 (VHR) and 2021/01700-2 (GT). MAL is supported by the U.S. Department of Energy grant DE-SC0020651 and acknowledges support of the Fulbright Commission of Brazil.
CS is supported by the U.S. Department of Energy under Award No. DE-SC0021969. CS acknowledges support from a DOE Office of Science Early Career Award.
JT acknowledges CNPQ bolsa de produtividade 309174/2020-1.
GT acknowledges CNPQ bolsa de produtividade 306152/2020-7 as well as grant BPN/ULM/2021/1/00039 from the Polish National Agency for Academic Exchange.
WMS is supported by FAPESP projects 22/11842-1, 21/01670-6 and 18/24720-6.

\end{acknowledgments}

\nocite{*}

\bibliography{apssamp}

\end{document}